\documentclass[10pt,journal,compsoc]{IEEEtran}
\usepackage{cite}
\newcommand{\myname}{ALP}

\newcommand{\squishlist}{
 \begin{list}{$\circ$}
  { \setlength{\itemsep}{0pt}
     \setlength{\parsep}{0pt}
     \setlength{\topsep}{3pt}
     \setlength{\partopsep}{0pt}
     \setlength{\leftmargin}{1em}
     \setlength{\labelwidth}{1em}
     \setlength{\labelsep}{0.5em} } }

\newcommand{\squishend}{
  \end{list}  }

\usepackage{fancyhdr}
\usepackage[normalem]{ulem}
\usepackage[hyphens]{url}
\usepackage[keeplastbox]{flushend}

\usepackage{setspace}
\usepackage{graphicx}
\usepackage{verbatim}
\usepackage{amsmath}
\usepackage{algorithmicx}
\usepackage{algpseudocode}
\usepackage{soul}
\usepackage{framed}
\usepackage[utf8]{inputenc}
 
\usepackage{listings}
\usepackage{xcolor}
 
\definecolor{codegreen}{rgb}{0,0.6,0}
\definecolor{codegray}{rgb}{0.5,0.5,0.5}
\definecolor{codepurple}{rgb}{0.58,0,0.82}
\definecolor{backcolour}{rgb}{0.95,0.95,0.92}
 
\lstdefinestyle{mystyle}{
    backgroundcolor=\color{white},   
    commentstyle=\color{codegreen},
    keywordstyle=\color{magenta},
    numberstyle=\tiny\color{codegray},
    stringstyle=\color{codepurple},
    basicstyle=\ttfamily\footnotesize,
    breakatwhitespace=false,         
    breaklines=true,                 
    captionpos=b,                    
    keepspaces=true,                 
    numbers=left,                    
    numbersep=5pt,                  
    showspaces=false,                
    showstringspaces=false,
    showtabs=false,                  
    tabsize=2
}
 
\usepackage[ruled,linesnumbered]{algorithm2e}

\usepackage{fancyhdr}
\usepackage{booktabs}
\usepackage{float}
\usepackage[normalem]{ulem}

\usepackage{breqn}

\usepackage{tikz}
\usepackage{xcolor}
\newcommand*\circled[1]{\tikz[baseline=(char.base)]{\node[shape=circle,fill,inner sep=0.5pt] (char) {\textcolor{white}{#1}};}}
\definecolor{denim}{rgb}{0.08, 0.38, 0.74}
\definecolor{azure(colorwheel)}{rgb}{0.0, 0.5, 1.0}
\definecolor{green(pigment)}{rgb}{0.0, 0.65, 0.31}
\definecolor{darkmagenta}{rgb}{0.55, 0.0, 0.55}
\definecolor{royalblue(web)}{rgb}{0.25, 0.41, 0.88}
\definecolor{ao(english)}{rgb}{0.0, 0.5, 0.0}

\newcommand{\rev}[1]{\textcolor{black}{#1}}

\usepackage[colorinlistoftodos,prependcaption,textsize=scriptsize]{todonotes}
\usepackage{dblfloatfix}    %

\usepackage{caption}
\usepackage{subcaption}

\sethlcolor {white}
\newcommand{\revtetc}[1]{\hl{#1}}

\ifCLASSINFOpdf
\else
\fi

\begin{document}
\title{\huge ALP: Alleviating CPU-Memory Data Movement Overheads\\in Memory-Centric Systems}

\author{Nika Mansouri Ghiasi,
        Nandita Vijaykumar,
        Geraldo F. Oliveira,
        Lois Orosa,
        Ivan Fernandez, \\
        Mohammad Sadrosadati,
        Konstantinos Kanellopoulos,
        Nastaran Hajinazar,
        Juan G\'{o}mez Luna,
        Onur Mutlu%
\IEEEcompsocitemizethanks{\IEEEcompsocthanksitem Nika Mansouri Ghiasi, Geraldo F. Oliveira, Lois Orosa, Mohammad Sadrosadati, Konstantinos Kanellopoulos, Nastaran Hajinazar, Juan G\'{o}mez Luna, and Onur Mutlu are with the Department of Information Technology and Electrical Engineering (D-ITET), ETH Zurich, 8092 Zürich, Switzerland.
\IEEEcompsocthanksitem Nandita Vijaykumar is with the Department of Computer Science, University of Toronto, Toronto, ON M5S 2B1, Canada.
\IEEEcompsocthanksitem Ivan Fernandez is with the Department of Computer Architecture, University of Malaga, 29016 Málaga, Spain.}%
}

\markboth{}
{Shell \MakeLowercase{\textit{et al.}}: ALP: Alleviating Host-Memory Data Movement Overheads\\in Memory-Centric Systems}

\IEEEtitleabstractindextext{%
\begin{abstract}
Recent advances in memory technology have enabled near-data processing (NDP) to tackle main memory bottlenecks in modern systems. 
Prior works partition applications into segments (e.g., instructions, loops, functions) and execute memory-bound segments of the applications on NDP computation units, while mapping the cache-friendly application segments to host CPU cores that access a deeper cache hierarchy. 
Partitioning applications between NDP and host cores causes inter-segment data movement overhead,  
which is the overhead from moving data generated from one segment and used in the consecutive segments. This overhead can be large if the segments map to cores in different parts of the system (i.e., host and NDP). 
Prior works take two approaches to the inter-segment data movement overhead when partitioning applications between NDP and host cores.
The first class of works maps segments to NDP or host cores based on the properties of each segment, neglecting the performance impact of the inter-segment data movement. Such partitioning techniques suffer from inter-segment data movement overhead.
The second class of works maps segments to host or NDP cores based on the overall memory bandwidth savings of each segment (which depends on the memory bandwidth savings within each segment and the inter-segment data movement overhead between other segments). These works do not offload each segment to the best-fitting core if they incur high inter-segment data movement overhead.  Therefore these works miss some of  the potential NDP performance benefits.
We show that mapping each segment (here basic block) to its best-fitting core based on the properties of each segment,  assuming no inter-segment data movement, can provide substantial performance benefits. However, we show that the inter-segment data movement reduces this benefit significantly.

To this end, we introduce ALP, a new programmer-transparent technique to leverage the performance benefits of NDP by
\emph{alleviating} the performance impact of inter-segment data movement between host and memory and  enabling efficient partitioning of applications between host and NDP cores.
\myname{} alleviates the inter-segment data movement overhead  by \emph{proactively and accurately} transferring the required data between the segments mapped on host and NDP cores.
This is based on the key observation that the instructions that generate the inter-segment data stay the same across different executions of a program on different input sets. 
\myname{} uses a compiler pass to identify these instructions and uses specialized hardware support to transfer data between the host and NDP cores at runtime.
Using both the compiler and runtime information, \myname{}  efficiently maps application segments to either host or NDP cores considering 1) the properties of each segment, 2) the inter-segment data movement overhead between different segments, and 3) whether this inter-segment data movement overhead can be alleviated proactively and in a timely manner. 
We evaluate ALP across a wide range of workloads and show on average 54.3\% and 45.4\% speedup compared to executing the application only on the host CPU or only the NDP cores, respectively.

\end{abstract}

}

\maketitle

\IEEEdisplaynontitleabstractindextext

\IEEEpeerreviewmaketitle

\IEEEraisesectionheading{\section{Introduction}\label{sec:introduction}}

\IEEEPARstart{N}{ear} 
data processing (NDP) paradigm improves overall system performance by alleviating the main memory bottlenecks~\cite{giannoula2022sparsep,gomez2022benchmarking,lee2016simultaneous, ahn2015scalable, nai2017graphpim, boroumand2018google, lazypim, top-pim, gao2016hrl, kim2018grim, drumond2017mondrian, RVU, NIM, PEI, gao2017tetris, Kim2016, gu2016leveraging, HBM, HMC2, boroumand2019conda, hsieh2016transparent, cali2020genasm,Sparse_MM_LiM, NDC_ISPASS_2014, farmahini2015nda,loh2013processing,pattnaik2016scheduling,akin2015data, hsieh2016accelerating,babarinsa2015jafar,lee2015bssync, devaux2019true,
Chi2016, Shafiee2016, seshadri2017ambit, seshadri2019dram, li2017drisa, seshadri2013rowclone, seshadri2016processing, deng2018dracc, xin2020elp2im, song2018graphr, song2017pipelayer,gao2019computedram, eckert2018neural, aga2017compute,dualitycache, fernandez2020natsa, hajinazarsimdram,syncron,boroumand2022polynesia,boroumand2021mitigating,amiraliphd,oliveira2017generic,kim2019d,kim2018dram,besta2021sisa,ferreira2021pluto,seshadri2016buddy,boroumand2017lazypim,kim2017grim,ghose2018enabling,seshadri2018rowclone,mutlu2019processing,mutlu2019enabling,ghose2019workload,olgun2021quactrng}.
While the  cores in modern systems are provided with deep and large cache hierarchies, NDP computation units suffer from lack of such an advantage due to their limited area and thermal budget~\cite{gao2016hrl, skarlatos2016snatch, tsai2018adaptive}.
Accordingly, prior works partition applications into segments (e.g., instructions, loops, functions) and execute memory-bound segments of the applications on NDP computation units, and map the cache-friendly application segments to host CPU cores that access a deeper cache hierarchy.

If not done properly, partitioning an application's code into NDP-friendly and CPU-friendly segments can result in a large volume of inter-segment data movement (i.e., data generated from one segment and used in other segments). When the segments map to the computation units on host and NDP systems, the data movement between the segments, in turn, translates to data movement overhead between the host and the NDP units and amortizes parts of the performance benefits of NDP. 
Prior works take two approaches to inter-segment data movement when partitioning applications between the host and NDP computation units.
The first class of works maps segments to the host or the NDP computation units based on the characteristics of each segment by considering the memory access behavior of each segment \emph{individually} \cite{oliveira2021damov, gao2015practical, boroumand2019conda}. Such works offload the memory-bound application segments  to the NDP computation units, and keep the more cache-friendly segments in the host CPU cores. 
Since these approaches consider the memory access behavior of each segment individually and isolated from the other segments, they suffer from inter-segment data movement overhead between the host cores and NDP computation units. 
The second class of works maps application segments to the host or NDP computation units based on the overall memory bandwidth savings of each segment, which depends on the memory bandwidth savings within each segment and the inter-segment data movement overhead between other segments~\cite{Nai2017, kim2017toward, hsieh2016transparent}. These works do not offload each segment to the best-fitting core if they incur high inter-segment data movement overhead.  Therefore, these works suffer from missing some of the potential NDP performance benefits.

To our knowledge, no prior work alleviates the cost of inter-segment data movement.
We show that while mapping each segment to its best fitting computation unit\footnote{As our NDP architecture, we consider a 3D-stacked memory with cores in its logic layer (called NDP cores), accessing memory with higher bandwidth and lower latency compared to the host CPU cores.} in the host or the NDP side provides significant benefits (on average 26.8\% and up to 44.1\% better than execution only on the host or NDP computation units), the inter-segment data movement overhead significantly reduces this potential and can even lead to slowdown compared to running the application only on the host computation units (on average 9.5\% and up to 56.3\% slower).

\textbf{Our goal} in this work is to alleviate the impact of inter-segment data movement to enable efficient partitioning of applications between NDP and host computation units.
To this end, we propose \textbf{ALP}, a programmer-transparent hardware-software cooperative mechanism that \underline{Al}leviates data movement between different segments when \underline{P}artitioning applications between NDP and host computation units. The key idea of \myname{} is to alleviate the inter-segment data movement overhead by \emph{proactively and accurately} transferring the required data between the segments that are mapped to the host and NDP computation units. This is based on the key observation that the instructions that generate the inter-segment data remain the same across different executions of a program on different input sets~\cite{marshaling}.

Quantifying and alleviating the overhead of inter-segment data movement while partitioning applications between the NDP and host units is challenging since they require 1)~identifying the inter-segment data that would cause performance overhead during partitioning, 2)~transferring the inter-segment data to the unit that will execute the next code segment before the next segment starts, and 3)~mapping segments between the NDP and host CPU units based on both  the segment's internal characteristics and the resulting inter-segment data movement (considering the timeliness of proactive inter-segment data transfers). Jointly considering these factors is challenging since they are impacted by the complex interaction between many features of the application, the input data, and the underlying architecture.

\myname{} leverages compile-time and runtime information and operates in three steps. We assume a system with cores in the host side connected to a 3D-stacked memory with cores on its logic layer (i.e., NDP cores). NDP cores access memory with higher bandwidth and lower latency compared to the host CPU cores.
In the first step, the compiler identifies the segments that the data movement between them could potentially reduce  performance if they map to different NDP/host cores. \myname{} marks these segments as \emph{tightly-connected segments.}

In the second step, using compile-time profiling, \myname{} finds the instructions that generate the inter-segment data in the tightly-connected segments~\cite{marshaling}. 
Then, \myname{} identifies clusters of tightly-integrated segments in which the inter-segment data can be proactively transferred from the generator segment to the producer segment in a timely manner. To do so, \myname{} identifies the cases where the time of transferring the inter-segment data can be fully hidden by other operations performed in the segments.
We identify the cases in which the time for transferring the data written by these instructions can be overlapped with the execution of other instructions in the segments. 
This way, during the application's runtime, \myname{}'s hardware can proactively transfer the inter-segment data to the core that consumes it, while hiding and eliminating the data movement overhead of this transfer. Using proactive data transfers, \myname{} enables starting the execution of the next segments of the applications as soon as their inter-segment data arrives. For example, instead of making the NDP computation units wait until the host writes back all the inter-segment data located in its large caches, \myname{} enables the NDP  computation units to start execution as soon as the parts of the inter-segment data arrives.

In the third step, during runtime,  \myname{} 
collects information regarding input data size, cache sizes, cache miss rates, and Instructions per Cycle (IPC) of the segments. \myname{} adds this information to the compile-time information about the tightly-connected segments collected in the first two steps.
By collectively considering these factors, \myname{} efficiently incorporates the information regarding the inter-segment data movement overhead in making partitioning decisions. \myname{} can be tuned to enrich various partitioning techniques by alleviating their inter-segment data movement overhead.

We evaluate \myname{} across workloads from various domains (e.g., graph processing, graphics, machine learning, bioinformatics, and high-performance computing). For workloads whose data movement overhead can be alleviated via proactive data transfers, \myname{} achieves almost all the potential performance benefits of mapping each segment of the application to its best-fitting core, assuming no inter-segment data movement overhead. \myname{} performs  on average 54.3\% better than execution only on the host CPU cores and 45.4\% better than execution only on the NDP cores. For workloads whose data cannot be proactively transferred, \myname{} does not incur any performance overhead by executing all segments in the NDP or host cores. \myname{} incurs a modest area overhead of 1.25KB and significantly improves the energy consumption.

We make the following contributions in this work:

\squishlist
    \item We identify and characterize a critical aspect of efficiently leveraging NDP: the impact of inter-segment data movement overhead between the NDP and CPU cores when the application is partitioned between them. We show that data movement overheads can significantly diminish the potential performance benefits of NDP. 
    \item We propose \myname{}, a programmer-transparent mechanism that alleviates the performance impact of data movement when partitions applications between the NDP and host CPU cores. \myname{} identifies the application segments that would incur high inter-segment data movement overhead during partition\revtetc{ing}, and alleviates this overhead by proactively and accurately transferring data between the segments. 
    \item \myname{} orchestrates the compile-time and runtime information about the inter-segment data movement overhead. \myname{} factors in  the characteristics of each segment, the estimated inter-segment data movement overhead, and  the timeliness of proactive data transfers during partitioning.
\squishend
\section{Background and Motivation}
\label{sec:background}

\subsection{Baseline Architecture}

The baseline system we assume in this work consists of a host CPU and a 3D-stacked memory module that supports processing data on the computation units in the logic layer. 
The logic layer and memory layers are connected using through-silicon vias (TSVs) \cite{lau2011overview}, which provide lower latency and significantly higher bandwidth than a traditional off-chip interconnection between main memory and the host CPU cores~\cite{ahn2015scalable, smc_sim, lazypim, lee2016simultaneous, NIM, NDC_ISPASS_2014,RVU,top-pim, boroumand2019conda, boroumand2018google,gao2017tetris, kim2018grim,cali2020genasm, fernandez2020natsa}. 
The host CPU and the NDP logic layer employ similar out-of-order (OoO) cores with different cache hierarchies. The host cores use a conventional three-level cache hierarchy, while the NDP cores use a single-level private cache. In this section, we model the same computation units in both host and NDP systems to decouple the effect of computation capabilities from memory hierarchy and data movement. We show the effects of different core types in Section~\ref{sec:results}. Section \ref{sec:evaluation} shows the details of our system organization and evaluation methodology.

\vspace{-0.5em}
\subsection{The Effect of Inter-Segment Data Movement}
\label{sec:motivation}

In this section, we show the performance impact of inter-segment data movement as a result of code partitioning between NDP and CPU cores. Applications can have different characteristics across different segments (e.g., basic blocks, loops, functions) \cite{gao2015practical, boroumand2019conda}. 
Segments of applications that suffer from main memory bottleneck take advantage of NDP execution, while more cache-friendly application segments take advantage of host CPU cores that access a deeper cache hierarchy. Therefore, executing the whole application on the host or NDP cores without partitioning  leads to missing opportunities of mapping each segment to the core it finds most beneficial.
Partitioning applications between NDP and host cores causes inter-segment data movement overhead (i.e., overhead from moving data generated from one segment and used in the consecutive segments). This overhead can be large if the segments map to cores in different systems (i.e., host and NDP).

Prior works take two approaches to the inter-segment data movement overhead when partitioning applications between NDP and host cores.
The first class of works~\cite{oliveira2021damov, gao2015practical, boroumand2019conda} maps segments to NDP or host cores based on architecture suitability of each segment. Such partitioning techniques suffer from inter-segment data movement overhead. 
The second class of works~\cite{Nai2017, kim2017toward, hsieh2016transparent} maps segments to host or NDP cores based on the overall memory bandwidth saving of each segment (which depends on the memory bandwidth saving within each segment and the inter-segment data movement overhead between other segments). These works conservatively do not offload the segments that would take advantage of NDP cores, but incur high inter-segment data movement overhead, missing some of  potential NDP performance benefits.

Through an idealized study, we show the potential benefit of NDP and how the performance impact of inter-segment data movement overhead on NDP benefits. To do so, we map each segment of the application to host or NDP systems based on the architectural suitability of each segment. We consider each basic block\footnote{We choose this because we find individual instructions to be too fine-grained for our NDP cores. This study can be performed at other granularities too.} as a segment and measure the execution time of each segment on an NDP core and on a CPU core to find out the best-fitting system (host or NDP) for each segment.  
Based on this oracle information, we map each segment to the core on which it performs best and compute the execution time of the programs \emph{with} and \emph{without} considering the performance impact of the inter-segment data movement overhead between the blocks.

\begin{figure}[h]
        \centering
        \includegraphics[width=0.9\linewidth]{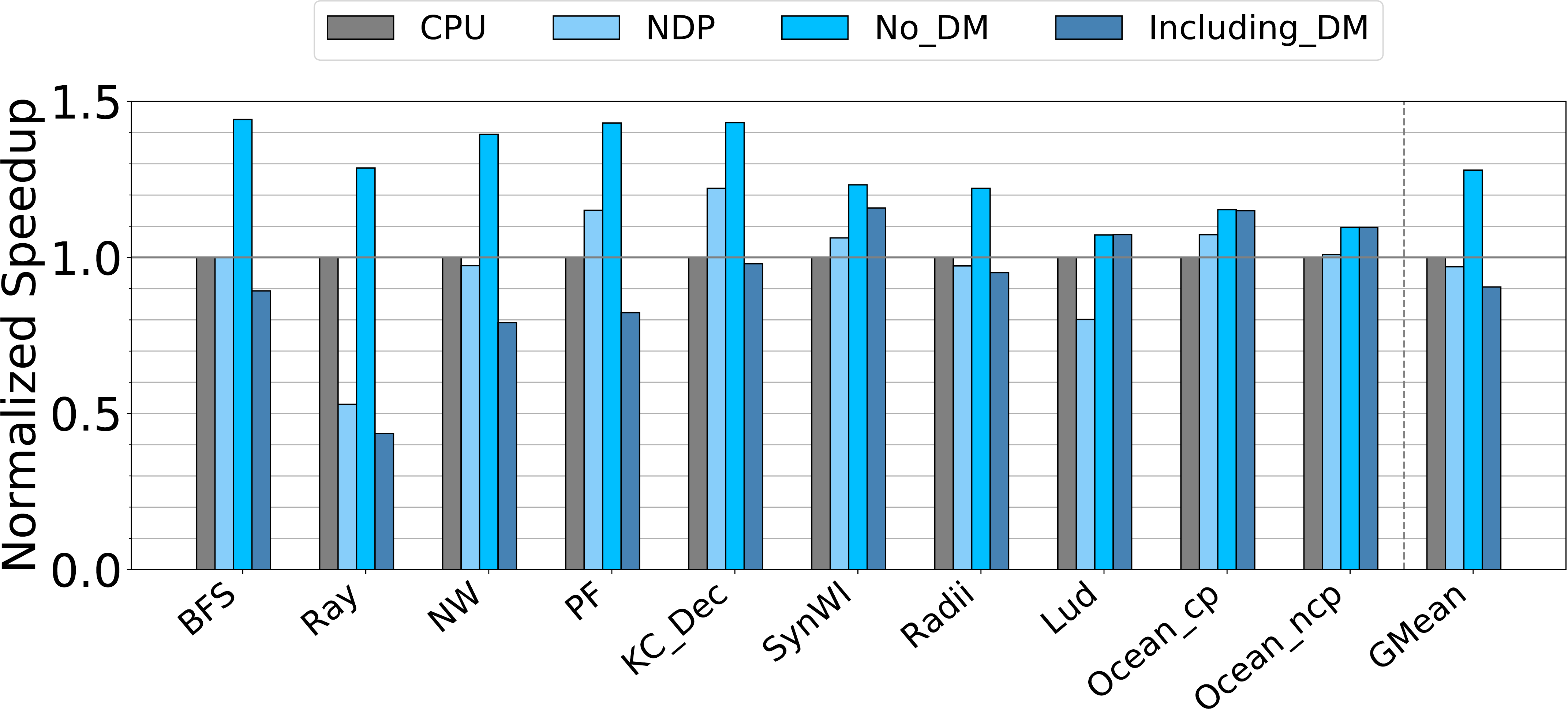}
        \caption{Performance effect of inter-segment data movement overhead.}
        \label{fig:motivation}
\end{figure}

Figure \ref{fig:motivation} shows the speedup of 1) executing  all segments of the application on the host CPU cores (\texttt{CPU}), 2) executing all the application on the NDP cores (\texttt{NDP}), 3) partitioning the application based on the architectural suitability of each segment with \emph{zero} data movement cost (\texttt{No\_DM}), 4) partitioning the application based on the architectural suitability of each segment and with the cost of data movement included (\texttt{Including\_DM}). The speedup values are normalized to the host CPU core's performance.
We make two observations based on this figure. First, \texttt{No\_DM} performs  on average 26.8\% (maximum 44.1\%) better than the best average performance of only NDP or only CPU execution. Second, with \texttt{Including\_DM}, the average speedup drops to 9.5\% (worst case 56.3\%) worse than \texttt{CPU}. 

\vspace{-2em}

\subsection{Goal}
 
Based on these observations, we conclude that even though the potential benefits of partitioning the applications to NDP and CPU cores in the absence of data movement overhead is very high, we see significant performance loss when we consider the  effect of data movement in making offloading decisions. We emphasize that our baseline system in this study is already equipped with prefetching (Table \ref{table_parameters}). However, as we see in Figure \ref{fig:motivation}, prefetching is not acting effectively in alleviating the data movement overhead because the access pattern of the data moved between the CPU and NDP cores due to code partitioning are typically irregular and non-repetitive. \textbf{Our goal} is to alleviate the impact of inter-segment data movement to enable efficient partitioning of applications between NDP and CPU cores

\rev{Prior works on partitioning \cite{austin1992dynamic, tseng2008achieving, jeffrey2015scalable} on heterogeneous core architectures do not study partitioning in the context of NDP and do not consider the asymmetry in the memory hierarchy. This leaves significant challenges to address in the NDP context.
First, performance and energy overhead of communication between NDP and CPU is very high due to \textit{off-chip} communication. Since the goal of NDP is reducing the overhead of data movement, this extra communication can amortize the potential benefits of NDP. This calls for a timely and proactive technique for addressing data movement issues in NDP. 
Second, prior works that propose techniques to alleviate data movement cost in a heterogeneous architecture ~\cite{marshaling} assume \textit{fixed and known} partitioning between the segments. In our scenario, we statically do not know where each segment maps.
Third, software or compiler-assisted prefetching \cite{intercoreSIGPLAN,ainsworth} techniques execute next to the code  executing in a different core, and therefore, do not transfer the data proactively. This cannot be \textit{timely} enough for NDP scenarios.
Due to these factors,  problem space of partitioning is very complex in this case because we need to consider (1) the advantage of NDP/CPU execution\revtetc{,} (2) significantly more critical data movement cost\revtetc{, and} (3) the potential for proactive data transfer.}

\vspace{-1.3em}
\section{ALP}
\label{sec:mechanism}

This section describes the three steps of \myname{}. In the first step (Section~\ref{sec:clustering}), during compile time, \myname{} detects the segments of the applications that can have high inter-segment data movement overhead. In the second step (Section~\ref{sec:marshal}), during compile time, \myname{} marks the instructions that generate the inter-segment data. In the third step (Section~\ref{sec:runtime}), during runtime, \myname{} incorporates the information about input data and the underlying architecture with the information collected during compile time and 1) performs proactive data transfer and 2) partitions applications between the host and NDP cores.

\vspace{-1em}
\subsection{Identifying Tightly-Connected Segments}
\label{sec:clustering}

The goal of this section is to identify the segments that the cost of data movement between them might amortize the cost of partitioning them. We refer to these segments as \emph{tightly-connected segments}. After detecting the tightly connected segments, next steps of \myname{} try to reduce the overhead of inter-segment data between these segments.
Listing \ref{listing:basic_marshal} shows an example of two tightly connected segments, assuming each loop is one segment. In the first loop, the application accesses several input arrays with random indices and generates \texttt{out}, which \revtetc{is} the inter-segment data between these two loops. $out$ is then re-used by the next nested loop for \texttt{n\_reuse} iterations. The random index \texttt{rand\_idx1} parameter in loop~1 is used to model random accesses to the data. Loop~1 accesses many data structures with random access patterns and maps better to NDP cores. However, if \texttt{n\_reuse} is high enough, and \texttt{out} is larger than what would fit in the smaller NDP caches, loop~2 will map better to the host CPU cores with larger caches. However, if we consider this mapping, the cost of transferring \texttt{out} to the host CPU amortizes some of the partitioning benefits. 

\begin{minipage}{\linewidth}
 \vspace{0.2cm}
\begin{lstlisting}[language=C++, caption=Synthetic workload for data transfer., label=listing:basic_marshal, basicstyle=\scriptsize]
for (uint64_t i = 0; i < N; i++){
    //accessing large vectors with random indices
    data = in1[rand_idx1] + in2[rand_idx1] + in3[rand_idx1];
    out.push_back(data);
}
for (uint64_t i = 0; i < n_reuse; i++){ 
    for(uint64_t j = 0; j < N; j++){
        //reusing the output of the previous segment
        out[j] = f(out[rand_idx2], i);
    }    
}
\end{lstlisting}
\end{minipage}

\begin{figure*}[b]
\centering
\includegraphics[width=15cm]{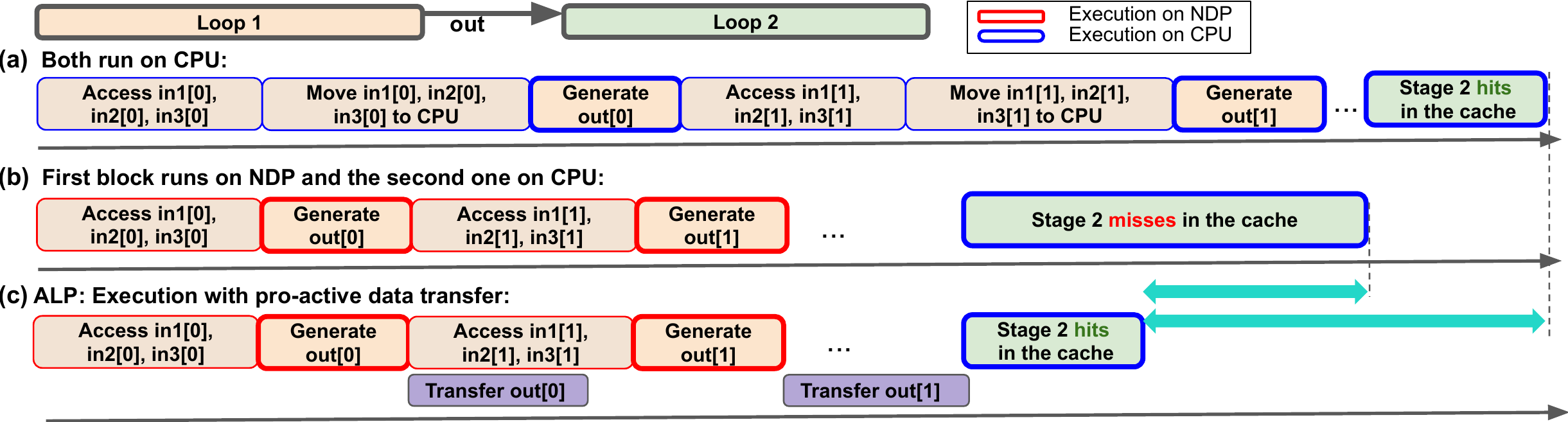}
\caption{Timeline of a workload on (a) CPU, (b) NDP, (c) with proactive data transfer (the indices are not the vector indices. Number $i$ inside the brackets refers to the $ith$ cache line accessed through the execution.)\vspace{-0.3cm}}
\label{fig:basic_marshal}
\end{figure*}

The first step of \myname{} leverages compiler's assistance to detect if two segments are tightly-connect by calculating their \texttt{connectivity}. The connectivity between the two segments depends on the ratio of the inter-segment data over all the data that both segments consume and produce. Thus,  the \texttt{connectivity} between two segments can be modeled as follows:

\begin{equation} \label{eq:connect}
\footnotesize
connectivity=\max\left(\frac{inter\_segment\_data}{reg\_in1+reg\_out1}, \frac{inter\_segment\_data}{reg\_in2+reg\_out2}\right),
\end{equation}

where $reg\_in1$ and $reg\_out1$ are the number of live registers moving in and out of the first segment respectively. \revtetc{$reg\_in2$} and \revtetc{$reg\_out2$} are the number of registers moving in and out of the second segment. \revtetc{$inter\_segment\_data$} is the number of the overlapping registers in $reg\_out1$ and $reg\_in2$ sets, which refers to the live registers that pass from one segment to the other. The liveness analysis of the compiler provides information regarding the live registers.
If  $connectivity$ exceeds an architecture-dependent threshold, the mechanism marks the two segments as tightly-connected. This threshold depends on multiple architectural features that determine whether the overhead of inter-segment data movement outweighs the benefits of partitioning the segments between NDP and host cores. These architectural features can affect the inter-segment data movement overhead and the execution time of segments on NDP or host cores. Such features are 1) the latency and bandwidth of the off-chip link between main memory and host system, 2) the internal latency and bandwidth of main memory, and 3) the latency, bandwidth, and size of NDP and host caches, and 4) NDP and host processor core features, such as their frequency and issue width. 

This threshold is determined by a one-time offline profiling since, for a given system, this architecture-dependent threshold does not change. This threshold is calibrated by profiling a wide range of application segments with inter-segment data on a given system. We choose this threshold conservatively to mark the tightly-connected segments. ALP’s runtime phase further considers application-dependent and runtime information to decide whether two tightly-connected segments can be partitioned between NDP and host cores.

\begin{figure}
     \centering
     \begin{subfigure}{0.5\linewidth}
         \centering
         \includegraphics[width=0.9\linewidth]{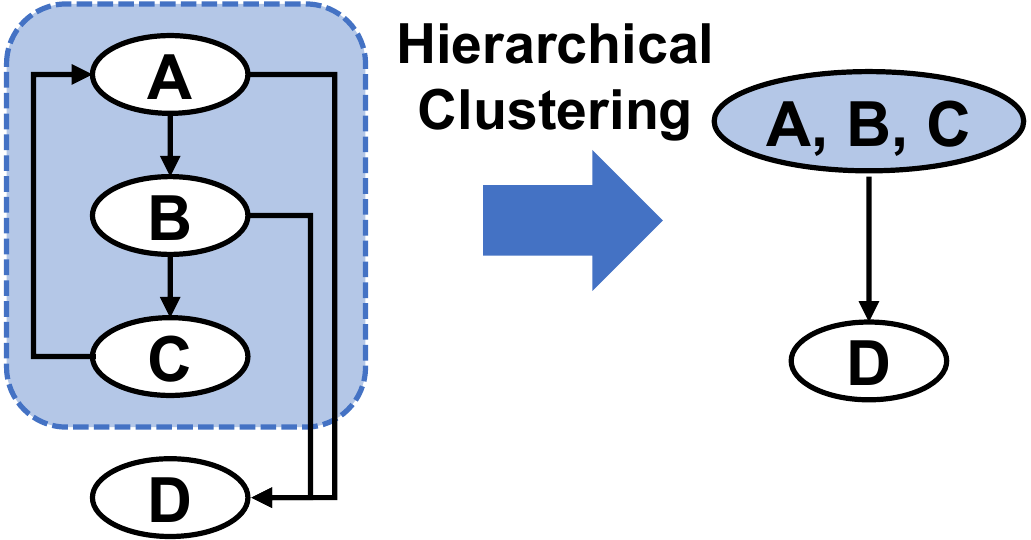}
         \caption{Loops}
         \label{fig:cluster-h-1}
     \end{subfigure}\hfill%
     \begin{subfigure}{0.5\linewidth}
         \centering
         \includegraphics[width=0.9\linewidth]{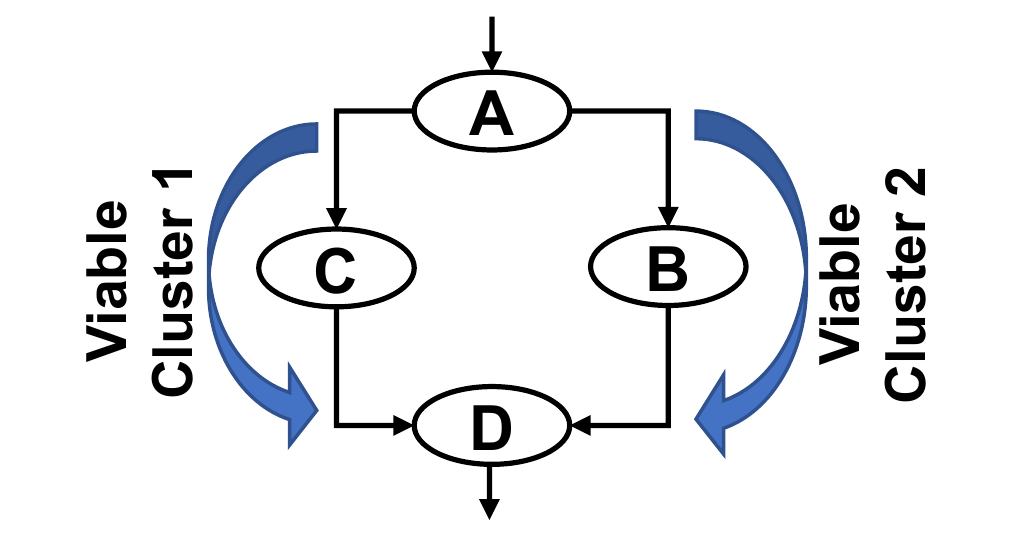}
         \caption{If-else statements}
         \label{fig:cluster-h-2}
     \end{subfigure}%
        \caption{Example of control flow divergence.}
        \label{fig:cluster-h}
\end{figure}

We calculate connectivity between application segments iteratively to find the segments of the code that have high data movement between each other. For example, in the control flow graph in Figure \ref{fig:cluster-h-1}, if segments A, B, and C form a cluster of tightly-connected segments, they might also form a larger cluster of tightly-connected segments with D. To model the data movement, we calculate the size of the inter\_segment data between the aggregated cluster (composed of segments A, B, and C) and segment D. Figure \ref{fig:cluster-h-2} shows an example of how \myname{} handles the control flow divergence using an if-else statement. We analyze both sides of the branch and mark the segments with a large amount of inter-segment data movement in each side. In case the connectivity between the segments in \textit{either} side is high, we mark the source and destination node of the control flow (A and D) and all intermediate segments as tightly-connected segments.

At the end of this stage, all the segments of the program are clustered with their tightly-connected segments. The data movement between the segments within a cluster might significantly reduce performance if the segments map to different NDP and CPU cores.
Section \ref{sec:implementation} explains the implementation details of how \myname{} passes this clustering information to its subsequent phases.

\subsection{Data Movement Alleviation}
\label{sec:marshal}

In this section, we explain how \myname{} alleviates the inter-segment data movement overhead between the highly connected segments. %

\subsubsection{Basic Data Transfer}
\label{sec:basic_marshal}

To illustrate our approach behind alleviating the inter-segment data movement overhead, we show the execution timeline of the synthetic workload in Listing~\ref{listing:basic_marshal} in three different cases in Figure \ref{fig:basic_marshal}. As mentioned in Section~\ref{sec:clustering}, assuming no inter-segment data movement overhead, loop~1 would ideally map best to the NDP cores and loop~2 would map best to the host CPU cores.
In case (a), both segments execute on the host CPU core. By the time loop~2 starts executing, \texttt{out} is present in the host caches and further accesses to it from loop~2 hit in the cache. Despite the efficient use of host CPU caches for accessing \texttt{out}, loop~1 suffers from memory bottleneck when running on the host CPU cores.
In case (b), the first segment executes in the NDP core, whereas the second segment executes in the CPU core. In this case, loop~1's memory bottleneck gets alleviated via NDP execution. However, in loop~2, all accesses to $out$ miss in the CPU caches and incur significant  data movement overhead from main memory. 
In case (c), we show \myname{}'s approach to reducing the inter-segment data movement overhead  which is enabled by proactively transferring the inter-segment data to the next segment, as soon as it is produced. Therefore, when loop~2 executes in the host CPU cores, it finds its needed data in host caches. 

Reducing the performance overhead of inter-segment data movement using this proactive data transfer approach can be possible if the time for transferring the data can be mostly overlapped with other operations. This means there should be more instructions between the instruction that \textit{generates} the inter-segment data and the instruction that \textit{consumes} the data. For example, in Figure \ref{fig:basic_marshal}, transferring $out[0]$ is overlapped with accessing $in1[1]$, $in[1]$, and $in3[1]$.

\subsubsection{Data Transfer with Concurrent Execution}
\label{sec:concurrent_marshal}
\begin{figure}[!t]
        \centering
        \includegraphics[width=\linewidth]{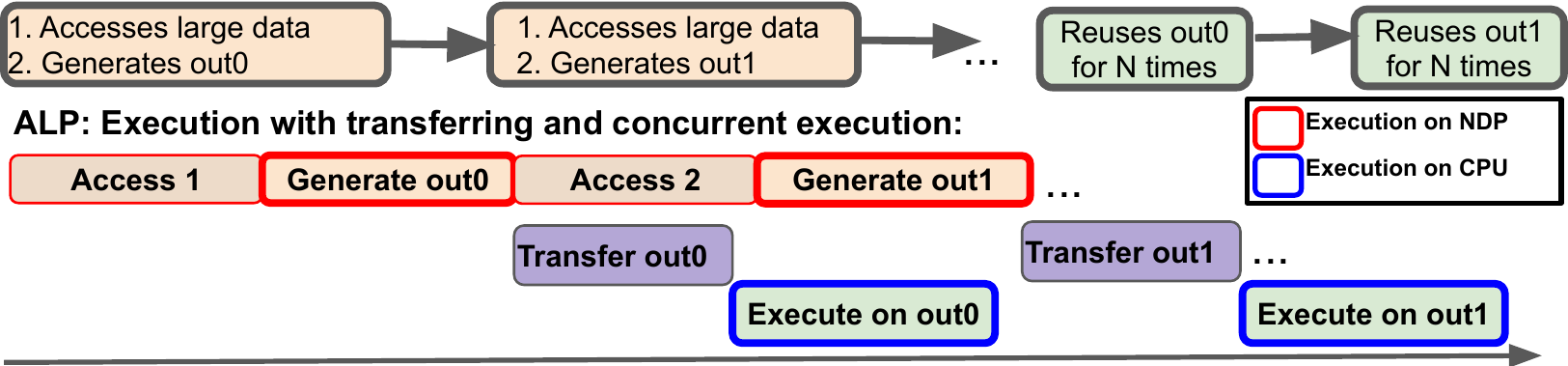}
        \caption{Execution timeline with transferring inter-segment data and concurrent execution of the segments. \vspace{-0.4cm} }
        \label{fig:online_marshal}
\end{figure}

In some workloads, some producer and consumer segments access the inter-segment data with the same access pattern.  In such cases, after the proactive data transfer of each cache-line of inter-segment data, the next segment of the application  can start execution on that data. This way, the segment that generates the inter-segment data elements can keep working on one core and the segment that consumes this data can run on another core, increasing the concurrency. Figure \ref{fig:online_marshal} shows an example execution timeline for this case. 
To detect this case, the compiler checks if (1) the first segment generates (writes) the data that second segment reads and (2) if the two segments access this data with the same access pattern.

\subsubsection{Detecting Inter-Segment Data}
\label{sec:detect_marshal}
In this section, we explain how \myname{} detects the inter-segment data that needs to be proactively transferred between the two tightly-integrated segments. 

In most programs, the instructions that generate the inter-segment data are the same across different executions of the program for different input datasets~\cite{marshaling}. Therefore, we leverage Data Marshalling technique~\cite{marshaling} to identify the instructions that generate the inter-segment data through profiling the application as shown in Algorithm~\ref{alg:profile_marshal}. 
This algorithm performs analysis on each two tightly-connected segments identified in the first step of \myname{} (Section~\ref{sec:clustering}). 
For every memory access in the current segment, the algorithm checks if the instruction that wrote to this address is from the previous segment. 
In that case, the last writer instruction to this address from the previous segment is marked as a generator instruction (Lines 1 to 5). 
For each write in the current segment, the algorithm collects the memory addresses, and the Programmer Counter (PC) in the $current\_last\_writer\_list$. 
This way, we can check if they are the generator instructions for the next segment (Lines 6 to 8).
After the end of each segment, we empty the  $previous\_last\_writer\_list$ for the previous segment, and set the current segment's  $current\_last\_writer\_list$ to be previous segment's  list $previous\_last\_writer\_list$ (Lines 10 to 14).

\begin{algorithm}[h]
\scriptsize 
\SetKwInOut{Input}{Input}
\SetKwInOut{Output}{Output}
\SetAlgoLined
\KwResult{generator\_instruction\_list}
\Input{Address of the memory accesses, Instruction PCs, previous\_last\_writer\_list}
 
  \For{Every memory access in the current segment}
   {\If{(accessed cache-line is first read in current segment) \textbf{and} (the address is in the previous\_last\_writer\_list)}{Add the PC of the last writer instruction in previous\_last\_writer\_list to generator\_instruction\_list}
  \If{Memory access is store}{Add the address and the PC to current\_last\_writer\_list}}
  \For{Every new segment start}
  {Empty previous\_last\_writer\_list\\
  Make current\_last\_writer\_list to be previous\_last\_writer\_list\\
  Make an empty list for current\_last\_writer\_list}
 \caption{Detecting generator instructions}
 \label{alg:profile_marshal}
\end{algorithm}

The compiler performs this profiling and marks the generator instructions with a new instruction added to Instruction Set Architecture (ISA). When the program executes these instructions, if the two segments map to different cores, \myname{} proactively transfers the data from one core's cache to the cache of the core executing the next segment. Section \ref{sec:impl_marsh} provides more details about the ISA and hardware support for this step.

ALP can work on any compiler because it relies on basic compiler features, like liveness analysis available in off-the-shelf compilers. The data movement analysis is done before the register allocation, in the IR stage, with the code in the static single assignment form. The baseline context-sensitive interprocedural analysis is required to model the data movements across the program. We analyze the whole graph of the application. \myname{} can adopt other techniques for optimizing interprocedural analysis. We do not expect our proposed mechanisms to significantly increase the compile time because they build on top of the already existing steps of compilation, like liveness analysis. Any additional increase in the compilation time will also be amortized over many runs for compiled languages.

\subsection{Incorporating Runtime Information}
\label{sec:runtime}

The goal of this step is to 1)~collect the architecture-dependent and runtime information and 2)~together with the compile-time information (collected in the first two steps of \myname{}), assess and incorporate the impact of inter-segment data movement overhead during partitioning.

\subsubsection{Offloading Metric}
\label{sec:runtime_offload_metric}

In this section, we explain the metrics that guide \myname{} to map segments to the host CPU or NDP cores. \myname{} can adopt other offloading metrics that can better suit different NDP architectures.

When an \emph{offload unit} (i.e., a segment or a cluster of segments that need to run together on the same core) starts execution on a host CPU core, we measure these metrics over a small epoch of execution. If the ratio of the L1 cache misses to the ratio of the LLC misses is close to one, we offload the execution to an NDP core. The reason is in these scenarios, the large LLC does not efficiently serve more memory requests compared to the small L1 cache. Since the NDP cores also have a small L1 cache, but higher bandwidth connection to main memory, these segments will potentially take more advantage of NDP execution.
After the NDP offload happens, the NDP core also measures the IPC over an epoch of execution. In case the IPC of NDP execution is lower than what was measured in the host CPU core before offloading, the execution migrates back to the CPU core. 
Section \ref{sec:impl_runtime_structure} describes the implementation details of \emph{Runtime Table} that keeps track of the runtime information of different runtime units over different epochs of their execution.

\subsubsection{Offloading Segments with Potential for Data Movement Alleviation}
\label{sec:runtime_offload_marsh}

This phase of \myname{} incorporates the information about the input data and the architecture features for the cluster of segments with the potential for data movement alleviation. 
As discussed in Section~\ref{sec:basic_marshal}, the data movement alleviation technique through proactive data transfer is effective if the time for transferring the inter-segment data can be mostly overlapped with other operations. This means there should be more instructions between the instruction that \textit{generates} the inter-segment data and the instruction that \textit{consumes} the data. We refer to the segments with these features as segments with potential for data movement alleviation.

Based on the size of the inter-segment data and the NDP and host cache sizes, and the characteristics of each segment, \myname{} maps the segments to the host CPU or NDP cores in two scenarios.
First, if the size of the inter-segment data is too large to fit in the destination cache, the basic data transfer scheme (Section~\ref{sec:concurrent_marshal}) will not improve the performance. The reason is that after some point, the new arriving parts of the inter-segment data will evict older parts. However, in case we can transfer data with concurrent execution (Section~\ref{sec:concurrent_marshal}), the next segment uses the data when it arrives.  
Second, if the size of the inter-segment data is small (compared to the destination's cache size), we assume its cost of data movement will not affect the total execution time unless the segments happen more than once. Therefore, \myname{} can profile the tightly-connected segments over a few iterations and map them to the host CPU or NDP cores based on the offloading metrics defined in Section~\ref{sec:runtime_offload_metric}.

Algorithm \ref{alg:runtime_marsh} shows how the runtime phase of \myname{} considers these different scenarios and maps these segments to NDP or CPU cores. \rev{Lines 1 to 21 in this algorithm handle the case where the inter-segment data in \textit{each iteration} is smaller than the size of the CPU cache. \revtetc{L}ines 22 to 30 handle segments with large inter-segment data.}

\begin{algorithm}[h]

\scriptsize
\SetKwInOut{Input}{Input}
\SetKwInOut{Output}{Output}
\SetAlgoLined
\KwResult{The mapping of each segment within the cluster to NDP or CPU}
\Input{Begin and end PC address of the cluster \\ The size of the inter-segment data}
 
  \eIf{size of inter-segment data $<$ CPU cache size}
  {
   profile the segments within the cluster on initial iterations\;
   \eIf{The producer segment shows high memory intensity}
   {\eIf{The consumer segment shows low memory intensity}{MAP(producer,NDP)\;
   MAP(consumer, CPU)\;
   Transfer the inter-segment data\;}{MAP(producer, NDP)\;
   MAP(consumer,NDP)\;}}
   {\eIf{The consumer segment shows low memory intensity}{MAP(producer, CPU)\; 
   MAP(consumer, CPU)\;}{MAP(producer,CPU)\;
   MAP(consumer,NDP)\;
   Transfer the inter-segment data\;}}
   }{
   Profile the producer segment in CPU\;
   \eIf{The producer segment shows high memory intensity \textbf{and} transfer with concurrent execution mode}{MAP(producer, NDP)\;
   proactively transfer the inter-segment data to CPU\;}{MAP(producer, CPU)\;}
  }
 
 \caption{Offloading Segments with the Potential for Data Movement Alleviation}
 \label{alg:runtime_marsh}
\end{algorithm}

\subsubsection{Offloading Segments without the Potential for Data Movement Alleviation}
\label{sec:runtime_offload_tight}
This section explains how \myname{} maps the segments within the same cluster of tightly-connected segments to the NDP or CPU cores in case the segments do \emph{not} take advantage of the proactive data transfer technique (as described in Section~\ref{sec:basic_marshal}). 

Based on the size of the inter-segment data and the NDP and host cache sizes, and the characteristics of each segment, \myname{} maps the segments to the host CPU or NDP cores in two scenarios.
First, if the data moved from one segment to another is very large, the cache of the NDP or host CPU cores cannot capture the re-use of this data between the segments. Therefore, executing the tightly-connected segments of the cluster on the same core does \textbf{not} lead to higher performance. 
Second, if the data movement can be captured in the host  caches, \myname{} makes offloading decision for them as a single  unit. We call these segments \emph{inseparable segments}. 

Mapping inseparable segments is challenging because their optimal mapping depends on their own characteristics and the characteristics of their connected segments. 
We make the key observation that the important inseparable segments occur \textit{jointly and repeatedly}. 
The reason is that if these segments do not occur frequently, they either  (1)~take small amount of execution time, therefore they do not contribute to the overall performance, or (2)~they take large amount of time and pass large amount of data to each other which cannot be captured in caches for further re-use. 
Therefore, by leveraging the repeated nature of the inseparable segments, \myname{}'s runtime mechanism profiles the \textbf{aggregated behavior} of them over an epoch to make offloading decision for both segments in their following iterations.
\section{Implementation}
\label{sec:implementation}

\begin{figure}[h]
        \centering
        \includegraphics[width=\linewidth]{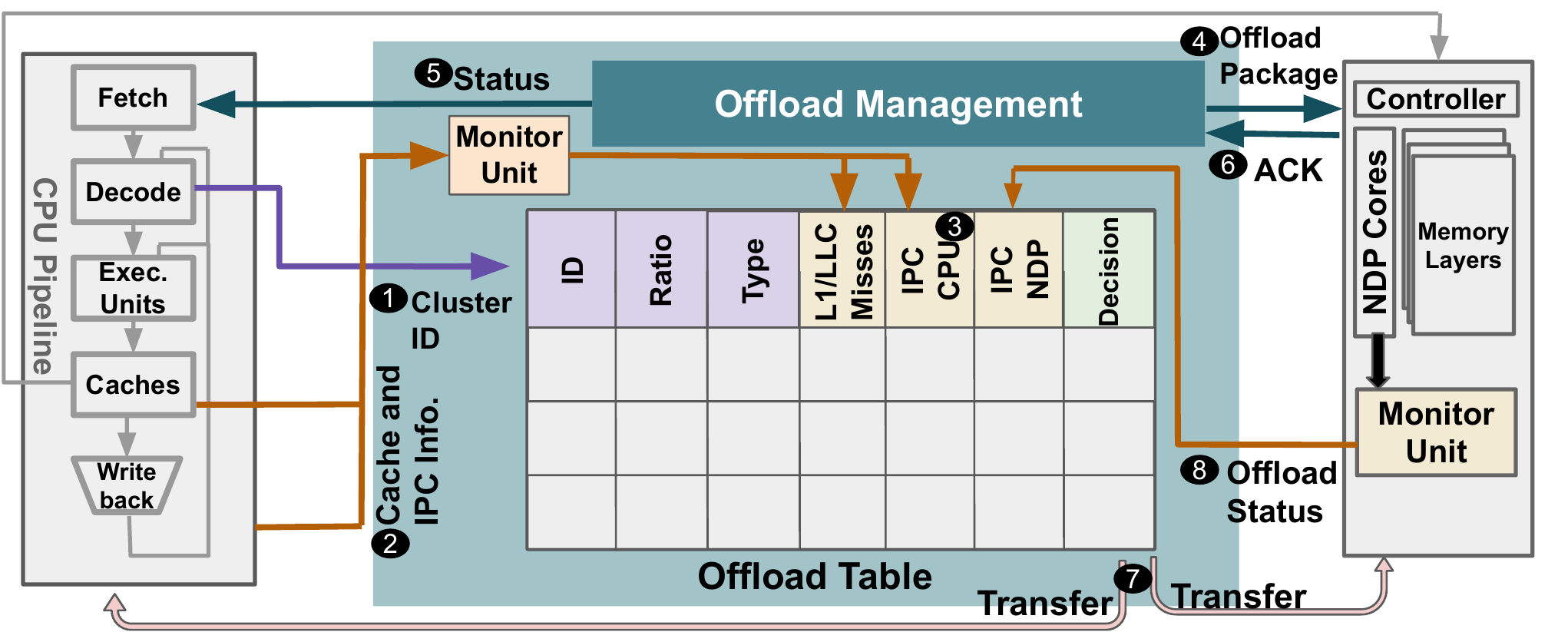}
        \caption{Overview of \myname{}'s hardware units. \vspace{-0.3cm}}
        \label{fig:hardware_units}
\end{figure}

In this section, we provide details on \myname{}'s implementation.
Figure \ref{fig:hardware_units} shows how \myname{}'s hardware units interface with the host CPU pipeline. We add an \emph{Offload Management Unit} that resides on the host chip, and is responsible for handling the offload to the NDP cores. 
The \emph{Monitor Units} in the NDP and CPU cores collect the necessary runtime information and populate the Offload Table. 
The Offload Table also holds the information that the compiler has gained about the segments of the code. Based on our analysis in Section \ref{sec:result_area}, this table can be accessed within one clock cycle. The static and dynamic information in the Offload Table is the basis for decision making of the Offload Management Unit for mapping the offload units (segments or a cluster of segments) to different cores. 
The hardware components of ALP (i.e., Offload Management Unit and  Monitor Units) are not in the critical path of any of the pipeline stages in the processor and reside next to the cores, and in each epoch,  receive information about the execution of the application segment on the host CPU cores and the NDP cores. Therefore, ALP does not change the processor cores pipeline’s depth and its frequency.
This section explains the functionality of these units in detail.

\subsection{Identifying Clusters}
\label{sec:impl_clusters}

In this section, we describe how the first step of \myname{} communicates the compile-time information about the segments (Section \ref{sec:clustering}) for the use of the next step of \myname{}.

\noindent\textbf{Code Annotation and ISA Support.} 
The compiler passes the information about the segments to the runtime mechanism by two new instructions that we introduce to the Instruction Set Architecture (ISA): {\fontfamily{qcr}\selectfont CLSTR.BEGIN} and {\fontfamily{qcr}\selectfont CLSTR.END}. %
These instructions identify respectively the beginning and the end of the cluster of the tightly-connected segments and include an identifier to each cluster. These instructions enable detecting when the candidate offload units start and ends, without any programmer involvement.

\noindent\textbf{Interface to the processor.} The compile-time information collected in the first step populates the \emph{Offload Table} located in the Offload Management Unit (Figure \ref{fig:hardware_units}) at the beginning of the program's execution, which is then accessed during runtime (during Phase 3). This table holds information about the segments, such as their ID, type (producer, consumer, or inseparable segments), and the relative inter-segment data ratios (i.e., the ratio of inter-segment live registers over the total live registers in the tightly-connected segments). During the execution, after decoding the instruction for {\fontfamily{qcr}\selectfont CLSTR.BEGIN},  \myname{} searches the table with the Cluster ID to retrieve the data about the clusters.

\subsection{Data Movement Overhead Alleviation}
\label{sec:impl_marsh}
This section explains how \myname{} performs the proactive data transfer between different segments of the application as described in Section~\ref{sec:marshal}.

\noindent\textbf{Code Annotation and ISA Support.}
After specifying the last writer of the inter-segment data, the compiler marks those instructions by adding a prefix called {\fontfamily{qcr}\selectfont TRANSFER} to them. This prefix indicates that the instruction's resulting data \rev{(inter-segment data)} needs to be transferred to the next segment as soon as it is generated. Therefore, when this instruction executes, the Offload Management Unit realizes that it has to transfer the inter-segment data to the next segment in case the segments within the cluster of tightly-integrated segments are mapped to different cores.  Using these instructions, \myname{} detects and transfers the inter-segment data in a timely \revtetc{manner,} regardless of the access pattern of the inter-segment data across different input sets and executions. The information collected during this step in the compile time also populates the Offload Table's entry, to indicate the segments types (producer of the inter-segment data or its consumer).

\noindent\textbf{Hardware Support.}
If the segments with a cluster of tightly-integrated segments are mapped to different cores, the runtime phase of \myname{} handles the data transfer through a modified MESI cache coherency protocol. If the data needs to transfer from the host CPU cores to the NDP cores, the runtime system issues this transfer request, and makes the inter-segment cache-lines invalid in the host caches, transfers the data to the NDP cache, and change its status as modified. If the data transfers from the NDP cores to the host CPU cores, the cache-line gets invalidated in the NDP cache and transfers to the host CPU caches with modified state.

\subsection{Runtime System}
\label{sec:impl_runtime_structure}

This section explains how the third phase of \myname{} makes the offloading decisions based on the static and dynamic information collected through its different steps.
Figure \ref{fig:hardware_units} shows the \textit{Offload Table}, which holds the information about different offload units, and the table's  interface to CPU and NDP cores. First, the compiler-generated data populates the table with clusters information.
The table also holds information about the cluster type to detect whether they have the potential for data transfer, and in case they do, it separates the producer and consumer segments in separate rows to profile them separately \circled{1}.
During the execution of the program, the Monitor Unit stores the information about the IPC and L1 and LLC cache miss rates of each offload unit in the Offload Table \circled{2}.
It also detects the absolute size of the inter-segment data based the actual data sizes and the ratio of the inter-segment data determined during compile-time.
After an epoch of execution, if the memory-intensity of an offload unit is high, the Monitor Unit records the Instruction per Cycle (IPC) of its execution so far in the Offload Table \circled{3}, 
and sends the offloading unit to the NDP cores. 
It also includes the required input live registers of the offload and its starting PC in the \emph{Offload Package} \circled{4}. 
The Offload Management Unit informs the host CPU core\revtetc{s} to stall \circled{5} until it receives an acknowledgment that the NDP execution has finished \circled{6}. 
Meanwhile, in case the producer and consumer blocks map to different cores, the Offload Management Unit issues the transfer signal to the respective coherency mechanism to move the inter-segment data from the producer to the consumer proactively \circled{7}. 
After each epoch of execution, the NDP core\revtetc{s} communicate the status of NDP execution by sending the IPC of the offload unit to the Offload Table. In case the IPC of the offload unit decreases from what has been measured in during its execution in the CPU, the offload unit transfers back to the host CPU for its remaining iterations \circled{8}.

\subsection{Design Considerations}

\subsubsection{Cache Coherence}
\label{sec:coherence}

In this work, we model a fine-grained coherency protocol between the NDP and  host CPU cores. The host LLC  is only inclusive of the host-side caches. \rev{Therefore back invalidation from LLC  only affects the CPU caches. When a cache-line is evicted from LLC, in case it is also present in an NDP cache, its status will change to Exclusive in the NDP cache.}
\rev{If there is any coherency-related data movement between NDP and CPU caches, we model its cost by adding the cost of memory stack's off-chip link to the baseline coherency overhead modeled by our simulator.}

Further optimization over the fine-grain coherency is possible by adopting more advanced coherency mechanisms such as \cite{boroumand2019conda}.  
\rev{The problem we address in this work is rather about the inter-segment data sharing between the segments, which exists even in the context of a single thread and is different from the coherency issue. 
Advanced coherency mechanisms do improve the performance of NDP execution, and their performance impact can be considered during \myname{}'s runtime.}

\subsubsection{Virtual Memory}

Translating virtual addresses to physical addresses when executing an application on the NDP side is challenging. \revtetc{If the NDP cores rely on the existing host-side translation mechanism, for every memory access, the NDP cores need to send a translation request to the host via low-bandwidth off-chip buses. This overhead can further increase if the translation requires page table walks that incur further data movement overhead between host and main memory. If the NDP cores naively duplicate TLB and page table walker, they can incur significant overhead due to 1) maintaining coherency with the host-side translation mechanism, and 2) additional area overhead.}  

Similar to previous works \cite{Tesseract}, in this work, we assume Direct Segments \cite{basu2013efficient} as our virtual memory model and interface the memory as a primary region. \revtetc{Each direct segment maps a large range of contiguous virtual memory addresses to contiguous physical addresses using base, limit and offset information.  If a virtual address  is between the base and limit, its corresponding physical address is simply translated as that virtual address plus the offset.
To support direct segment translation, the NDP cores require a small direct segment hardware (including registers to store base, limit, and offset values, and an adder) and need to receive base, limit, and offset values from the host at the beginning of NDP execution of each offloaded application segment. ALP can orthogonally adopt more advanced NDP-specific translation techniques} \cite{picorel2017near, ESMC_DATE_2015} \revtetc{for further performance benefits, which are beyond the scope of this work.
}

\subsection{\rev{Multiple NDP Stacks}}

ALP extends to a system with multiple NDP stacks assuming a baseline mapping \textit{between the stacks}. ALP addresses the problem of data sharing \textit{between segments in the NDP and host CPU cores}. After offloading a segment to NDP cores, ALP’s runtime structure (Section~\ref{sec:impl_runtime_structure}) monitors its execution on NDP cores.  If the NDP stacks have high data movement between each other and that leads to lower performance than host execution, ALP can identify that and rollback the execution to CPU in case that leads to higher performance.

\section{Evaluation Methodology}
\label{sec:evaluation}

\subsection{Experimental Setup}

We develop a system level simulator that accurately models the host CPU and NDP cores with the data movement between them. Our simulator uses ZSim \cite{sanchez2013zsim}  (https://github.com/s5z/zsim) to model the host and NDP cores and Ramulator~\cite{kim2016ramulator}   (https://github.com/CMU-SAFARI/ramulator) to model the 3D-stacked DRAM~\cite{oliveira2021damov,HMC2}. We modify ZSim and Ramulator so that the host CPU cores have lower bandwidth connections to DRAM, while the NDP cores have higher bandwidth connections to DRAM. 
We use Pin \cite{luk2005pin} to obtain information about the registers of the segments for phase 1 of \myname{}. To model the operation of phase 2, we develop a profiling tool based on algorithm \ref{alg:profile_marshal}. We model the hardware structures and the runtime analysis of phase 3 in ZSim.
We model a crossbar in our memory model in Ramulator to model the communication of data between different NDP cores in different vaults on the logic layer of the 3D-stacked system.
Our proposals can also be adopted to other  host + NDP architectures with asymmetric memory hierarchy properties.

Table~\ref{table_parameters} describes the system configuration we use to evaluate our proposal. Our system consists of x86 Out-of-Order (OoO) cores for both host and NDP sides.\footnote{When using different processors or Instruction Set Architectures (ISA), the compiler can also generate appropriate code based on our techniques.} The host and NDP cores have private L1  caches, but the host cores also leverage L2 and LLC which are only inclusive for the host. 
The NDP and host CPU cores run at the same 2.4~GHz frequency, with the goal of decoupling the effects of data movement and the memory hierarchy from the processing capabilities. We  demonstrate our  analysis on a single core CPU and single core NDP to isolate other effects (e.g., the interactions between the threads and sharing resources) from the inter-segment data movement overhead. We \revtetc{also} show the advantages of \myname{} with more cores (\revtetc{configuration with 8 NDP cores and 8 host CPU cores, and configuration with 32 NDP cores and 8 host CPU cores}).

\begin{table}[h]
    \centering
    \caption{Baseline, HMC and NDP configurations.}
    \label{table_parameters}
    \resizebox{\columnwidth}{!}{
    \begin{tabular}{@{} l @{}}
    \toprule
    \textbf{OoO Execution Cores}  @ 2.4~GHz, 32~nm; 4-wide out-of-order;   \\
        128-entry ROB; 32-entry LQ and 32-entry SQ; \\
        Branch predictor: Two-level GAs. 4,096~entry BTB; 1~branch per fetch;    \\
    \midrule
    \textbf{L1 Data + Inst. Cache} 32~KB, 8-way, 4-cycle; 64~B line; LRU policy; \\
    5/33 pJ per hit/miss \cite{energycache}\\
    \midrule
    \textbf{L2 Cache (only CPU)}  Private 256~KB, 8-way, 7-cycle; 64~B line; LRU policy; \\
        Prefetcher: Stream prefetcher with 16 entries;\\
        6/93 pJ per hit/miss \cite{energycache}\\
    \midrule
    \textbf{LLC Cache (only CPU)} Shared 8~MB, 16-way, 27-cycle; \\
        64~B line; LRU policy; Inclusive for CPU; MESI protocol; \\
        945/1904 pJ per hit/miss \cite{energycache}\\
    \midrule
    \textbf{3D-stacked DRAM} 4~GB, 32 vaults, 8 DRAM banks/vault; \\
        DRAM: CAS, RP, RCD, RAS and CWD latency (9-9-9-24-7 cycles);\\
        2 pJ/bit SerDes links \cite{kim2013memory}; 2 pJ/bit internal, 8 pJ/bit logic layer \cite{gao2015practical};\\ 
    \midrule
    \bottomrule
    \end{tabular}
 }
\end{table}

\subsection{Evaluated Applications}

Table \ref{table_workloads1} shows the list of the workloads we use in this work and their respective input sizes. We select a broad range of applications from popular benchmark suites and various domains. %
To demonstrate different partitioning scenarios, the selected applications include a wide range of memory-intensive and compute-intensive workloads. The \emph{common feature} of these workloads is that they have some segments that ideally would take advantage of execution on the NDP cores and some other segments that would take advantage of execution on the host CPU cores. Therefore, for each of these applications, the \textit{potential} performance benefit from partitioning the application between the NDP and host CPU cores is high.

\begin{table}[h]
\caption{Evaluated workloads and input sets.}
\label{table_workloads1}
\centering
\resizebox{\linewidth}{!}{%
\begin{tabular}{@{}cccc@{}}
\toprule
\textbf{Application} & \textbf{Benchmark Suite} & \textbf{Domain} & \textbf{Input Parameters} \\ \midrule
KCore-Decomposition & Ligra \cite{ligra} & Graph Processing & rMat 1M \cite{rmat} \\
Radii & Ligra \cite{ligra} & Graph Processing & rGnutella \cite{gnutella} \\
RayTrace & Parsec \cite{parsec} & Graphics & simlarge \\
Backpropagation & Rodinia \cite{rodinia} & Machine Learning & 524288 elements \\
Breadth-First Search & Rodinia \cite{rodinia} & Graph Processing & graph1MW \\
\rev{Breadth-First Search} & \rev{Ligra} \cite{ligra} & \rev{Graph Processing} & \rev{rMat 1M} \cite{rmat} \\
Needleman-Wunsch & Rodinia \cite{rodinia} & Bioinformatics & dimension 4096 penalty 10 \\
Particle Filter & Rodinia \cite{rodinia} & Statistics & x 128 y 128 z 10 np 400000 \\
Ocean (contiguous) & Splash-2 \cite{splash2} & High-Performance Computing & 514×514 grid \\
Ocean (non-contiguous) & Splash-2 \cite{splash2} & High-Performance Computing & 514×514 grid \\ \bottomrule
\end{tabular}%
}
\end{table}

To further analyze the workloads, we profile the applications to analyze their memory-bound behavior using Intel
VTune~\cite{vtune} on an Intel Xeon E3-1240 processor with 4 cores. Table \ref{table_workloads_vtune} shows the most memory bound function of each workload and the amount of time it takes in the whole application. Memory-bound measure refers to the ratio of cycles spent \textit{waiting for memory accesses} over the total execution time. 
We observe that all our selected workloads have functions that take a notable amount of execution time which are memory-bound. We conclude that these applications would ideally benefit from partitioning between the host and NDP cores.

\vspace{-0.2cm}
\begin{table}[h]
\footnotesize
\caption{\rev{Workload Characteristics.}}
\label{table_workloads_vtune}
\centering
\resizebox{\linewidth}{!}{%
{\begin{tabular}{@{}ccccc@{}}
\toprule
\textbf{Workload} & \textbf{Function} & \textbf{Time (\%)} & \textbf{Mem-bound (\%)} & \textbf{Mem. accesses} \\ \midrule
Kcore-decomposition            & edgeMapDense          & 52.7      & 53.82  & 2.6~GB\\
Radii                          & edgeMapDense          & 80.78     & 52.41  & 136~MB\\
RayTrace                       & {[}VTune format{]}    & 62.35     & 6.52   & 70.13~MB\\
Backpropagation                & bpnn\_adjust\_weights & 61.82     & 86.50  & 3.9~GB\\
Breadth-First Search (Rodinia) & BFSGraph              & 5.45      & 22.54  & 1.39~GB\\
Breadth-First Search (Ligra)   & edgeMapDense          & 30.86     & 34.08  & 2.5~GB\\
Needleman-Wunsch               & nw\_optimized         & 42.54     & 39.66  & 81~MB\\
Particle Filter                & {[}Vtune format{]}    & 3.99      & 2.70   & 3.09~MB\\
Ocean (contiguous)             & slave2                & 24.41     & 22.98  & 2.62~GB\\
Ocean (non-contiguous)         & slave2                & 16.00     & 21.96  & 1.23~GB\\ \bottomrule
\end{tabular}%
}
}
\end{table}

\section{Evaluation}
\label{sec:results}

In this section, we show the performance and energy benefits of \myname{} for various workloads. 
Throughout this section, \textbf{No\_DM} refers to partitioning the application based on the architectural suitability of each segment with \emph{zero} data movement cost, and \textbf{DM\_Included} refers to partitioning the application based on the architectural suitability of each segment with the cost of data movement included.

\subsection{Performance}
\label{sec:results_marshal}

In this section, we analyze the performance benefit of ALP,
compared to only host CPU, only NDP, DM\_Included and No\_DM execution for the workloads with the potential for data movement alleviation. As described in Section~\ref{sec:basic_marshal}, reducing the performance overhead of inter-segment data movement using the proactive data transfer approach can be possible if the time for transferring the data can be mostly overlapped and hidden with other operations. This means there should be more instructions between the instruction that \textit{generates} the inter-segment data and the instruction that \textit{consumes} the data. We show the performance benefits of \myname{} for workloads without the potential for proactive data transfer in Section~\ref{sec:results_tight}. Synth\_WL is the synthetic workload in Listing \ref{listing:basic_marshal} to demonstrate the proactive data transfer technique. 

\begin{figure}[h]
        \centering
        \includegraphics[width=\linewidth]{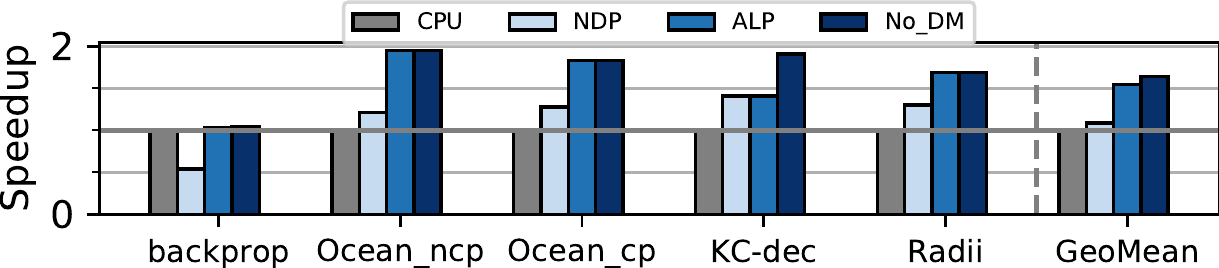}
        \caption{\rev{Performance benefits of \myname{}  with total 16 cores.}\vspace{-0.3cm}}
        \label{fig_8_8}
\end{figure}

\rev{Figure \ref{fig_8_8} shows the performance benefit of ALP with 8 NDP cores and 8 host CPU cores. We observe that ALP performs 54.3\% better than host-only and 45.4\% better than NDP-only execution.}
The memory-bound applications segments with high memory bandwidth requirements are able to issue more concurrent memory accesses in the NDP configuration with larger number of cores and larger available main memory bandwidth. The effectiveness of NDP execution improves in systems with larger number of cores because they take better advantage of the high main memory bandwidth available to the NDP cores. However, the application segments that take better advantage of larger caches take advantage of execution on the host cores. We conclude that \myname{} enables efficient partitioning of application segments between the host and NDP cores through efficient inter-segment data movement alleviation.

\begin{figure}[h]
        \centering
        \includegraphics[width=0.9\linewidth]{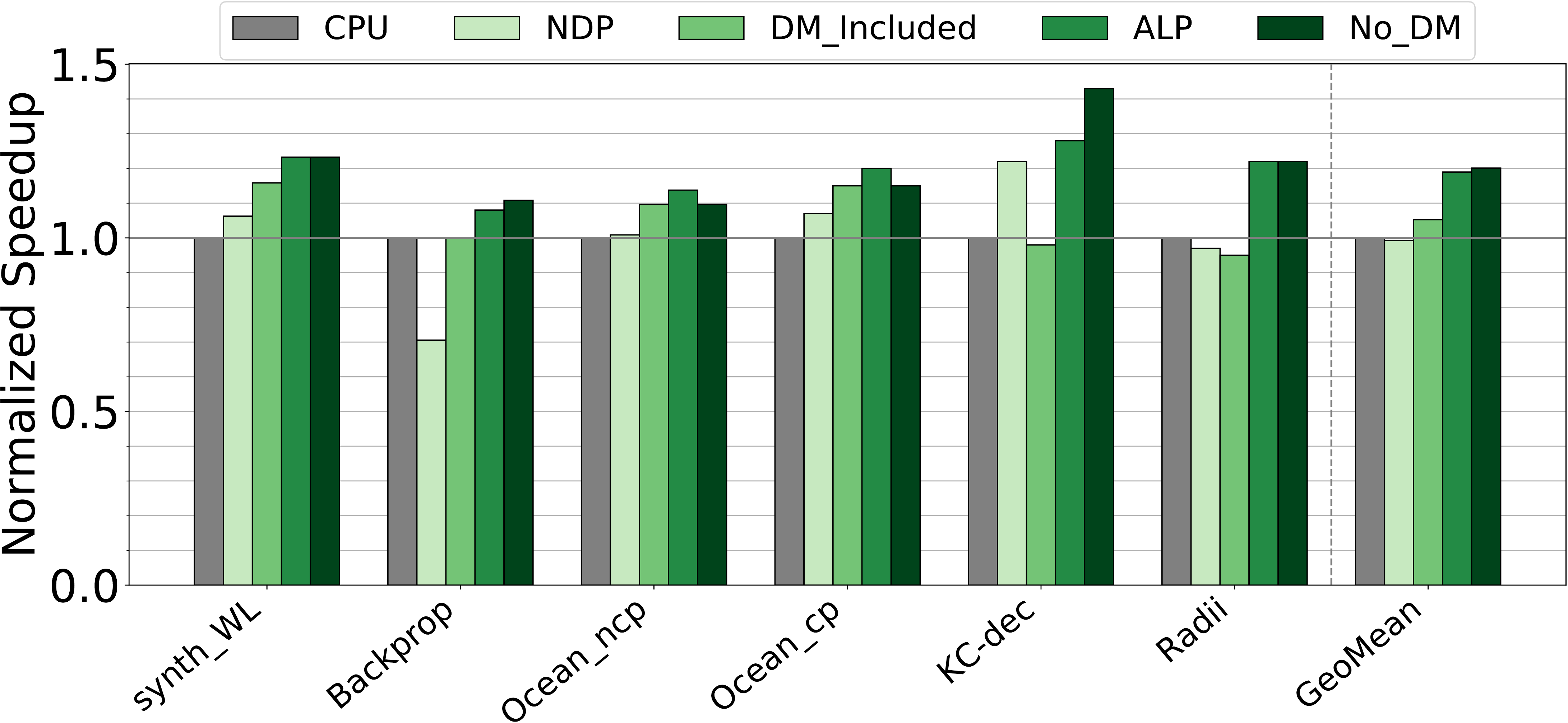}
        \caption{Performance benefits of \myname{} with data transfers. \vspace{-0.3cm}}
        \label{fig_marshal}
\end{figure}

Figure~\ref{fig_marshal} shows the performance benefits of \myname{} for single-core scenario to gain deeper understanding of how \myname{} alleviates inter-segment data movement overhead, by isolating other effects such as thread communication. Based on this figure, we make two observations. 
First, On average, ALP achieves almost all of the potential performance benefits of partitioning, achieving on average 18.9\% speedup over execution only on a host CPU core, and 19.7\% better than execution only an NDP core.
For most these workloads, \myname{}  achieves all of the potential benefits of partitioning because it can move the inter-segment data in a timely manner.
For some other workloads (ocean\_cp and ocean\_ncp), \myname{}  outperforms the \texttt{No\_DM} configuration  because the consumer segments can start the execution concurrently as soon as their required inter-segment data arrives (as described in Section~\ref{sec:concurrent_marshal}).
For some other workloads (Backprop, KC-dec), this technique enhances performance, but it does not reach the maximum possible performance because alleviating the data movement cost is only possible for some clusters. 
Second, \myname{} outperforms the performance of the DM\_Included case for all the workloads, even though in the DM\_Included case, each segment maps to the core on which it individually performs best. The reason is that when partitioning, \myname{} considers the  effect of the inter-segment data movement between the segments and alleviates its performance overhead.
We conclude that using both the compiler and runtime information, \myname{}  efficiently maps code segments to either host or NDP cores considering 1) the architectural suitability of each segment, 2) the inter-segment data movement overhead of each segment, and 3) whether this inter-segment data movement overhead can be alleviated proactively and in a timely manner.

\subsection{The Effect of Core Counts \rev{and Types}}

\begin{figure}[h]
        \centering
        \includegraphics[width=\linewidth]{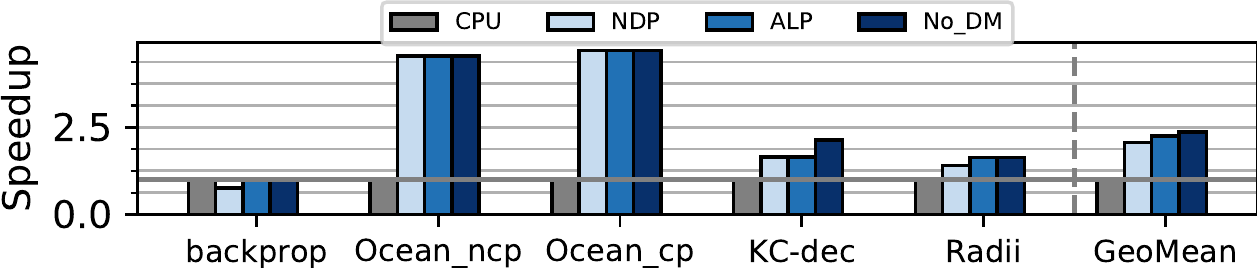}
        \caption{\rev{Performance benefits of \myname{} with 8 CPU cores and 32 in-order NDP cores.} \vspace{-0.3cm}}
        \label{fig_8_32}
\end{figure}

To study the effect of having different number and types of cores in host and NDP configurations, we study an NDP configuration with 32 in-order cores and a host configuration with 8 OoO cores. Figure \ref{fig_8_32} shows ALP's performance benefits in such a system.
We see that NDP benefits from the higher core count in this case. The runtime system of \myname{} (as described is Section \ref{sec:runtime}), using steps 1, 2, and 7, collects the IPC of the segments on both CPU and NDP cores. This way, it will detect the higher performance benefits of NDP and make the right decision for offloading segments accordingly. In this case, \myname{} performs $2.24~\times$ faster than execution only on the host CPU cores, and even 22\% faster than NDP-only execution. The reason is that although the NDP configuration has much larger number of cores in this case, there are still some application segments that take advantage of the larger cache hierarchy in the host.
We conclude that \myname{} adapts to different system configurations with various \rev{numbers and types of cores} by incorporating the architecture, input, and runtime information during the third phase.

\subsection{Energy}
\label{sec:result_multicore}

In this section, we show energy benefits of \myname{}.
The energy consumption of executing a segment of application on an NDP core is the sum of the energy spent on the cores, L1 NDP caches, and DRAM. The energy consumption of executing a segment of application on a host CPU core is the sum of the energy spent on the cores, L1, L2, and LLC CPU caches, off-chip links, and DRAM. ALP’s energy is the sum of the energy spent on the cores, L1, L2, and LLC CPU caches (for segments that access these caches), L1 NDP caches (for segments that access this cache), off-chip links (for data movement between NDP and CPU cores and for CPU memory accesses), and DRAM. The value of energy per access for each of these elements are listed in Table~\ref{table_parameters}.

\begin{figure}[b]
        \centering
        \includegraphics[width=0.9\linewidth]{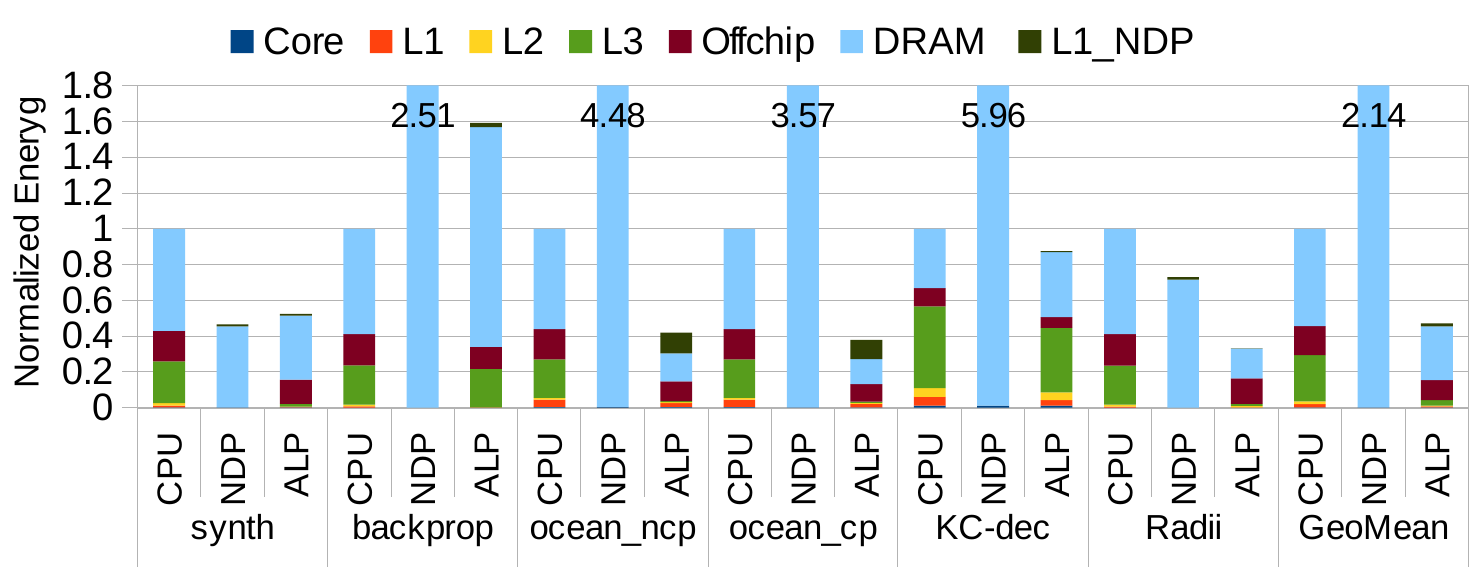}
        \caption{Energy consumption of ALP. \vspace{-0.3cm}}
        \label{fig_energy}
\end{figure}
Figure \ref{fig_energy} shows the energy consumption of only-NDP execution, only-host CPU execution, and \myname{}. We observe that \myname{} provides significant energy improvement over both NDP and CPU executions (4.5$\times$ and 2.12$\times$). The reason is that segments that map to a host CPU core take advantage of the large LLC and reduce the number of accesses that go to memory. They also capture the re-use between the segments that have high data movement between each other, avoiding extra off-chip data communication. Segments that map to an NDP core are those with random memory accesses which would have lead to high LLC miss rates. By executing these on NDP cores, they do not pay the extra cost of accessing the LLC and then subsequently bringing data from DRAM to the LLC via the off-chip links.

\subsection{Segments without Data Movement Alleviation}
\label{sec:results_tight}

\begin{figure}[h]
        \centering
        \includegraphics[width=\linewidth]{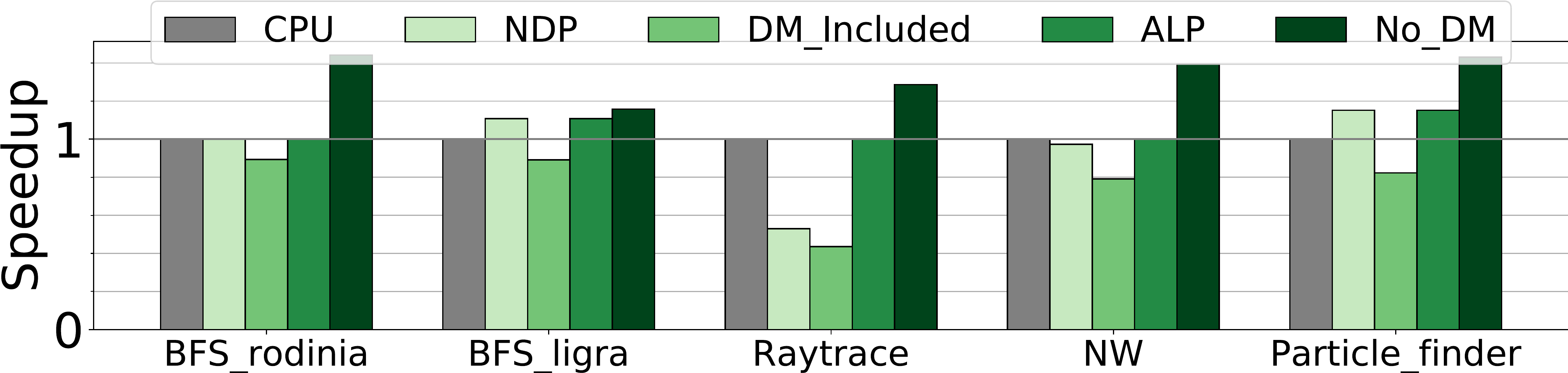}
        \caption{\rev{Performance benefit of \myname{} for inseparable segments.} \vspace{-0.3cm}}
        \label{fig:result_runtime}
\end{figure}

In this section, we present \myname{}'s performance benefit for the inseparable segments as described in Section \ref{sec:runtime_offload_tight}. These are the segments without TRANSFER instructions because the data transfer between them could not be overlapped with other instructions. 
Figure~\ref{fig:result_runtime}, we show the performance benefit of ALP for inseparable segments. We make two key observations. 
First, \myname{} performs on average 32.8\% better than mapping segments based on their individual characteristics (DM\_Included).
\myname{} avoids mapping these segments to different cores by considering the effect of inter-segment data movement. \myname{} profiles the \textit{aggregated behavior} of the segments over the epochs of execution and maps the inseparable segments to the core that they \textbf{collectively} find to be the most profitable candidate. 
This way, \myname{} avoids the performance loss that would have resulted from neglecting the inter-segment data movement overhead between these segments. 
Second, although Particle Finder is heavily compute-bound, we observe that it takes advantage of NDP execution. The reason is that the working set of this application is very small such that it can even fit in the small NDP caches. Therefore, \myname{}'s runtime system (Section~\ref{sec:runtime}) detects that the host-side LLC is not more efficient than L1, and offloads the application segments to the NDP cores accordingly. In this case, the workload performs better on the NDP cores because it does not spend extra time on the unnecessary L2 and LLC accesses in the host.

\subsection{Area Overhead}
\label{sec:result_area}

In this section, we determine the area overhead of \myname{} by calculating the size of the Offload Table ($Table\_size$) as follows:

\begin{dmath} 
\footnotesize
\label{eu_eqn}
Table\_Size = Row\_Count \times 
(Ratio + ID + Block\_Type + L1LLC\_Ratio, IPC \times 2 + Decision)
\end{dmath}

where the number of table rows ($Row\_Count$) is determined by the number of distinct offload units. In this work, we use 50 rows which is significantly more than the maximum number of offload units \myname{} extracts for the applications we studied. With 4 bits for representing the ratio of the inter-segment data ($Ratio$), 6 bits for block ID ($ID$), 2 bits for block types producer, consumer, or inseparable segments ($Block\_Type$), and 4 bits for ratio of cache misses ($L1LLC\_Ratio$) and 4 bits for considering IPC in in 16-level granularity ($IPC$), and one decision bit ($Decision$), the table size becomes 1.25~KB, which is significantly less than L1 cache size. Based on our CACTI \cite{chen2012cacti} simulations, a table of this size can be accessed within one clock cycle. In case other applications have large number of offload units, the rows of the Offload Table can be filled with an LRU policy.

\section{Related Work} 
\label{sec_related}

To our knowledge, \myname{} is the first programmer-transparent mechanism  to alleviate the inter-segment data movement overhead between the host and NDP cores by proactively transferring data between application segments. In this section, we discuss prior works that are related to different aspects of our work.

\subsection{Offloading Applications to NDP Computation Units}

Prior works take two approaches to inter-segment data movement when partitioning applications between the host and NDP computation units.
The first class of works maps segments to host or NDP based on the characteristics of each segment by considering the memory access behavior of each segment \emph{individually} \cite{oliveira2021damov, gao2015practical, boroumand2019conda}. Such works offload the memory-bound application segments  to the NDP computation units, and keep the more cache-friendly segments in the host CPU cores. 
For example, CoNDA \cite{boroumand2019conda}  assumes a programmer-annotated partitioning between NDP and CPU, and its goal is to enable efficient coherency between the partitions.  DAMOV~\cite{oliveira2021damov} identifies new insights about the different data movement bottlenecks and uses these insights to determine whether NDP or other data movement mitigation techniques are suitable for different applications. The key focus of these works does not target the problem of finding an efficient approach for partitioning the application to alleviate the overall data movement overhead.
Since these approaches consider the memory bottlenecks of each segment individually and isolated from the other segments, they suffer from inter-segment data movement overhead between the host cores and NDP execution units.

The second class of works maps application segments to the host or NDP computation units based on the overall memory bandwidth saving of each segment, which depends on the memory bandwidth saving within each segment and the inter-segment data movement overhead between other segments~\cite{Nai2017, hsieh2016transparent, kim2017toward, weipimprof, ahmed2019compiler, lee2001automatically}. If partitioning segments leads to high inter-segment data movement overhead, these works do not partition such segments to different cores that would be most beneficial for each segments. Therefore, as shown in Section~\ref{sec:motivation}, these works suffer from missing some of the potential benefits of partitioning. On the other hand, \myname{} proposes techniques for alleviating inter-segment data movement overhead to enable efficient application partitioning between NDP and host cores. \myname{} can be tuned to be adopted in different NDP proposals assuming different execution units in the logic layer.

\subsection{Co-Locating Computation and Data}

Prior work has studied placing data close to computation in different contexts. Hardware prefetchers \cite{sms,spp} do not properly detect the inter-segment data with irregular access patterns. Complex prefetchers \cite{imp,vlp} need long time to train over a large set of data, however, this cannot be timely for small inter-segment data or when the execution moves fast between CPU and NDP cores. Software prefetchers \cite{intercoreSIGPLAN,ainsworth} execute next to the code in the next segment, therefore, do not transfer the data proactively. Works on thread migration acceleration \cite{jeffrey} and caching techniques \cite{beckmann2015} are orthogonal to this work and can further improve \myname{}'s performance. 
Prior works on co-locating computation and data do not study inter-segment data movement in the context of systems with the host and NDP computation units and do not consider the asymmetry in the memory hierarchy. These factors leave significant challenges to address in the context of NDP. First, performance and energy overhead of communication between NDP and host computation units are very high due to off-chip communication. Since NDP's goal is reducing the overhead of data movement, this extra communication due to data sharing with the host can reduce the potential benefits of NDP. This work addresses these challenges in the context of NDP using compiler and hardware support.

Data Marshalling~\cite{marshaling} mitigates inter-core data misses in Staged-Execution models using proactive data transfers (marshalling).
While \myname{} also uses proactive data transfers as part of its second step, it is not known statically where each segment maps. This factor, along with the expensive off-chip communication in  NDP scenario, impose more challenges for efficient data transfer in the context of NDP and host systems.
Data Marshalling also does not address the problem of efficiently partitioning applications, and assumes the stages of the applications are already known.

Livia \cite{livia} proposes a new system architecture and programming model that co-locates tasks and data throughout the memory hierarchy with the goal of reducing the data movement. In \myname{}, we solve a different problem. We show that if two segments would ideally map to different NDP/CPU cores, the cost of data movement between them could amortize the benefits of partitioning. Therefore, by alleviating the data movement between them, we enable them to map to their ideal core.
\myname{} can be integrated in Livia's system to further improve its performance by alleviating inter-segment data movement between different parts of the application mapped in different computation units throughout the memory hierarchy. Tang et al. in \cite{tang2017data} propose a compiler algorithm that maps the computations to different NDP cores to reduce the distance-to-data on the on-chip network. \myname{}'s approach can further improve the performance for this proposal by applying proactive data transfer.

\section{Conclusion}
\label{sec:conclusion}

We identify and characterize an important aspect of NDP: the inter-segment data movement overhead between NDP and CPU cores when code is partitioned between them. We demonstrate that the inter-segment data movement overhead can significantly diminish the potential performance benefits from NDP. 
To fully leverage NDP, we introduce a programmer-transparent hardware-software cooperative mechanism, \myname{}, that (1)~considers and alleviates the performance impact of data movement and (2)~efficiently partitions applications between NDP and CPU, factoring in both architectural suitability and estimated data movement overhead.
Our analyses on a wide range of workloads show that \myname{} can achieve almost all the benefits of partitioning in workloads with the potential for proactive data transfer.

\ifCLASSOPTIONcaptionsoff
  \newpage
\fi

\bibliographystyle{IEEEtranS}
\bibliography{refs}

\begin{IEEEbiography}
    [{\includegraphics[width=1in,height=1.25in,clip, keepaspectratio]{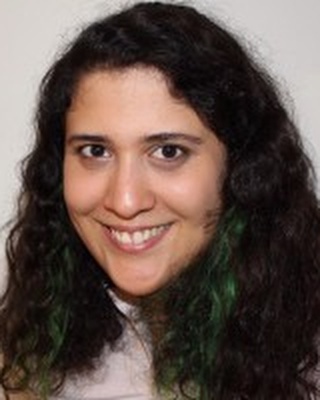}}]{Nika Mansouri Ghiasi} received the B.S. degree in Electrical Engineering from the University of Tehran, and the M.S. degree in Electrical Engineering from ETH Zürich. She is currently pursuing the Ph.D. degree at ETH Zürich, where she is advised by Onur Mutlu. Her research interests include emerging memory and processing technologies, near-data processing, storage systems, and bioinformatics.
\end{IEEEbiography}

\begin{IEEEbiography}
    [{\includegraphics[width=1in,height=1.25in,clip, keepaspectratio]{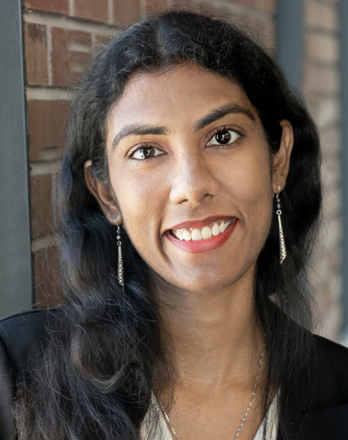}}]{Nandita Vijaykumar}  received the M.S. and Ph.D. degrees from Carnegie Mellon University, in 2019, where she was advised by Prof. Onur Mutlu and Prof. Phil Gibbons. She is currently an Assistant Professor with the Computer Science Department, University of Toronto, and the Department of Computer and Mathematical Sciences, University of Toronto Scarborough, and is affiliated with the Vector Institute. Before joining the University of Toronto, she was a research scientist with the Memory Architecture and Accelerator Laboratory, Intel Labs. In the past, she worked for AMD, Intel, Microsoft, and Nvidia. Her research interests include computer architecture, compilers, and systems with a focus on the interaction between programming models, systems, and architectures. Her current interests include the system-level and programming challenges of robotics, and large-scale machine learning. For more information, please visit her website at  \url{http://www.cs.toronto.edu/∼nandita/}.
\end{IEEEbiography}

\begin{IEEEbiography}
    [{\includegraphics[width=1in,height=1.25in,clip, keepaspectratio]{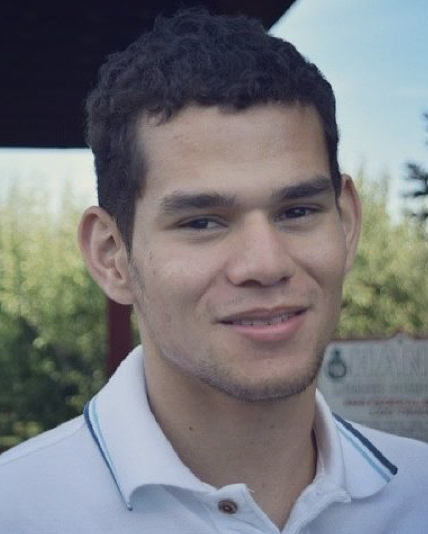}}]{Geraldo F. Oliveira} received the B.S. degree in computer science from the Federal University of Viçosa, Viçosa, Brazil, in 2015, and the M.S. degree in computer science from the Federal University of Rio Grande do Sul, Porto Alegre, Brazil, in 2017. He is currently pursuing the Ph.D. degree with ETH Zürich, Zürich, Switzerland, under the supervision of Onur Mutlu. His current research interests include system support for processing-in-memory and processing-using-memory architectures, data-centric accelerators for emerging applications, approximate computing, and emerging memory systems for consumer devices. He has several publications on these topics.
\end{IEEEbiography}

\begin{IEEEbiography}
    [{\includegraphics[width=1in,height=1.25in,clip, keepaspectratio]{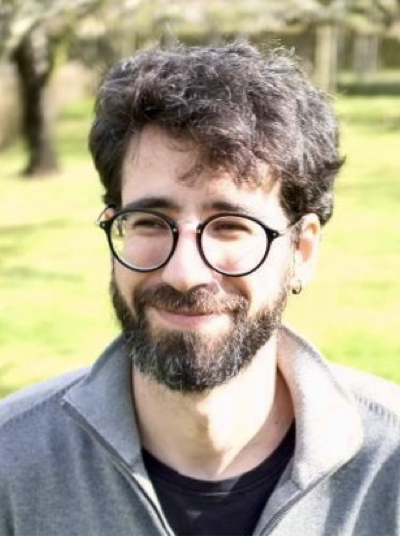}}]{Lois Orosa} is a senior researcher at SAFARI Research group at ETH Zürich, Switzerland. He received his BS and MS degrees in Telecommunication Engineering from the University of Vigo, Spain, his Ph.D. degree from the University of Santiago de Compostela, Spain, and he held a postDoc position in the University of Campinas, Brazil. He was a visiting researcher at multiple companies (IBM, Recore Systems, Xilinx and Huawei) and universities (UIUC and Universidade Nova de Lisboa). His current research interests are in computer architecture, hardware security, reliability, memory systems, and machine learning (ML) accelerators. For more information, please see his webpage at https://loisorosa.github.io/.
\end{IEEEbiography}

\begin{IEEEbiography}
    [{\includegraphics[width=1in,height=1.25in,clip, keepaspectratio]{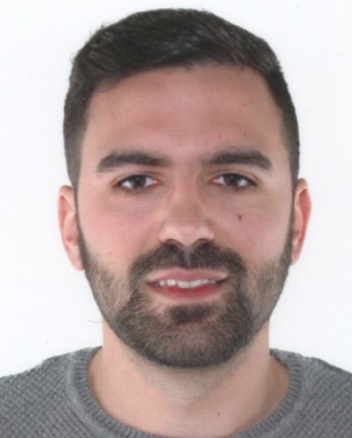}}]{Ivan Fernandez} received the B.S. degree in computer engineering and the M.S. degree in mechatronics engineering from the University of Malaga, in 2017 and 2018, respectively, where he is currently pursuing the Ph.D. degree. His current research interests include processing in memory, near-data processing, stacked memory architectures, high-performance computing, transprecision computing, and time series analysis.
\end{IEEEbiography}

\begin{IEEEbiography}
    [{\includegraphics[width=1in,height=1.25in,clip, keepaspectratio]{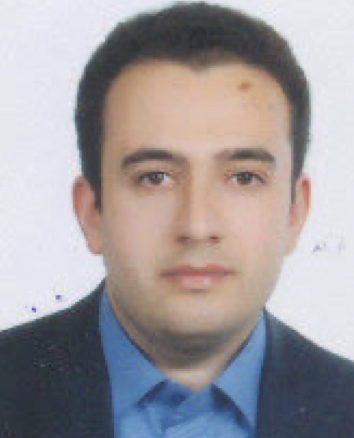}}]{Mohammad Sadrosadati} received the B.Sc., M.Sc., and Ph.D. degrees in computer engineering from Sharif University of Technology, Tehran, Iran, in 2012, 2014, and 2019, respectively. From April 2017 to April 2018, he spent one year as an Academic Guest at ETH Zurich, hosted by Prof. Onur Mutlu during his Ph.D. program. He is currently a Postdoctoral Researcher with ETH Zurich, working under the supervision of Prof. Onur Mutlu. His research interests include heterogeneous computing, processing-in-memory, memory systems, and interconnection networks. Due to his achievements and impact on improving the energy efficiency of GPUs, he received Khwarizmi Youth Award, one of the most prestigious awards, as the first laureate, in 2020, to honor and embolden him to keep taking even bigger steps in his research career.
\end{IEEEbiography}

\begin{IEEEbiography}
    [{\includegraphics[width=1in,height=1.25in,clip, keepaspectratio]{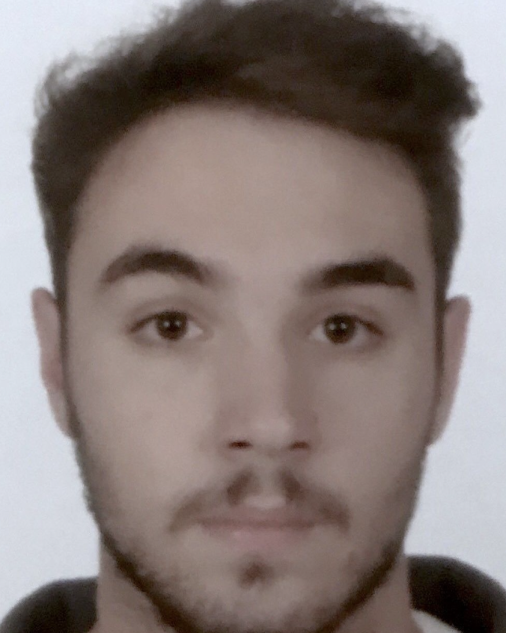}}]{Konstantinos Kanellopoulos} received the B.S. and M.S. degree in Computer Science from the National Technical University of Athens. He is currently pursuing the Ph.D. degree at ETH Zürich, where he is advised by Onur Mutlu. His research interests include hardware/software interfaces and hardware security.
\end{IEEEbiography}

\begin{IEEEbiography}
    [{\includegraphics[width=1in,height=1.25in,clip, keepaspectratio]{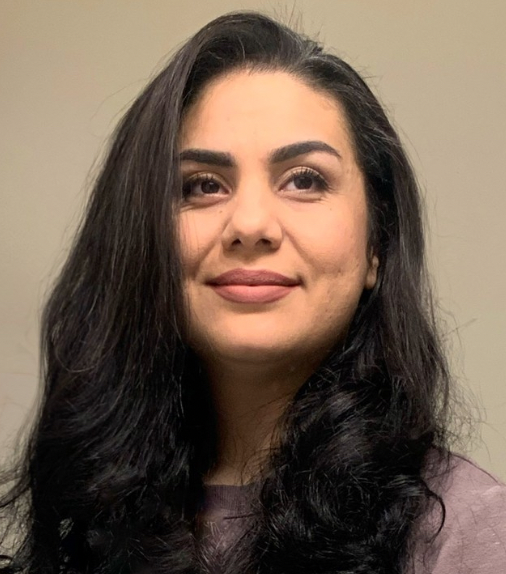}}]{Nastaran Hajinazar} is a Senior Researcher at ETH Zürich. Nastaran received her M.S. degree in computer hardware engineering from Sharif University of Technology, Iran, in 2011 and her Ph.D. degree in computer science from Simon Fraser University, British Columbia, Canada, in 2020. Her research incorporates several aspects of computer architecture with a significant focus on designing efficient high- performance computing systems, memory architectures, and intelligent memory management techniques.
\end{IEEEbiography}

\begin{IEEEbiography}
    [{\includegraphics[width=1in,height=1.25in,clip, keepaspectratio]{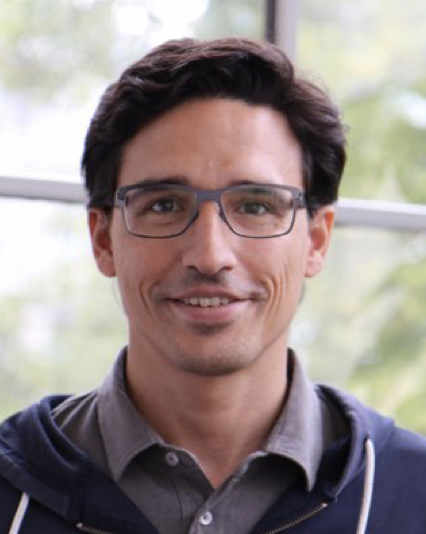}}]{Juan G\'{o}mez Luna} received the B.S. and M.S. degrees in telecommunication engineering from the University of Seville, Spain, in 2001, and the Ph.D. degree in computer science from the University of Córdoba, Spain, in 2012. From 2005 to 2017, he was a Faculty Member of the University of Córdoba. He is currently a Senior Researcher and a Lecturer with SAFARI Research Group, ETH Zürich. He is the lead author of PrIM (https://github.com/CMU-SAFARI/prim-benchmarks), the first publicly-available benchmark suite for a real-world processing-in-memory architecture, and Chai (https://github.com/chai- benchmarks/chai), a benchmark suite for heterogeneous systems with CPU/GPU/FPGA. His research interests include processing-in-memory, memory systems, heterogeneous computing, and hardware and software acceleration of medical imaging and bioinformatics.
\end{IEEEbiography}

\begin{IEEEbiography}
    [{\includegraphics[width=1in,height=1.25in,clip, keepaspectratio]{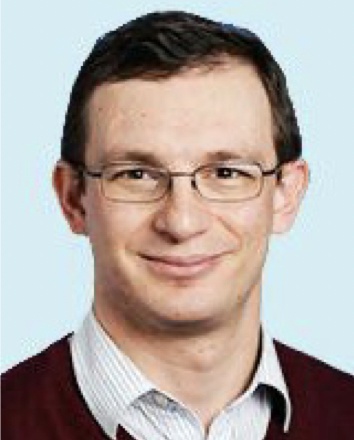}}]{Onur Mutlu (Fellow, IEEE)} received a B.S. degree in computer engineering and psychology from the University of Michigan, and the M.S. and Ph.D. degrees in electrical and computer engineering from the University of Texas at Austin. He is a Professor at ETH Zurich and a Faculty Member with Carnegie Mellon University, where he was Strecker Early Career Professor. He started the Computer Architecture Group at Microsoft Research and held various positions at Intel, AMD, VMware, and Google. His research interests include computer architecture, systems, hardware security, and bioinformatics. He is an ACM Fellow and an Elected Member of the Academy of Europe. He received the IEEE HPCA Test of Time Award, IEEE CS Edward J. McCluskey Award, ACM SIGARCH Maurice Wilkes Award, and faculty partnership awards from various companies, and a healthy number of best paper recognitions. More information in (https://people.inf.ethz.ch/omutlu/).
\end{IEEEbiography}

\end{document}